\documentclass[a4paper,useAMS,usenatbib]{mn2e}
\usepackage{times}
\usepackage{longtable}
\usepackage{url}
\usepackage{epsfig}
\usepackage{amsmath}
\usepackage{gensymb}
\usepackage{natbib}
\usepackage{graphicx}
\usepackage{aas_macros}
\usepackage{float}
\usepackage{amssymb}
\usepackage{amsmath}
\begin{document}

\title[Galaxy dust radiative transfer with \textsc{DART-Ray}]{Predicting the stellar and non-equilibrium dust emission spectra of 
high-resolution simulated galaxies with \textsc{DART-Ray}}

\author[Natale et al.]{\parbox{\textwidth}{Giovanni Natale$^{1,3}$, Cristina C. Popescu$^{1,2,3}$, Richard. J. Tuffs$^{3}$, 
Victor P. Debattista$^{1}$, J\"{o}rg Fischera$^3$, Meiert W. Grootes$^3$}\vspace{0.4cm}\\\\
\parbox{\textwidth}{$^{1}$Jeremiah Horrocks Institute, University of Central Lancashire, Preston, PR1 2HE, UK\\
$^{2}$The Astronomical Institute of the Romanian Academy, Str. Cutitul de Argint 5, Bucharest, Romania \\ 
$^{3}$Max Planck Institute f\"{u}r Kernphysik, Saupfercheckweg 1, D-69117 Heidelberg, Germany\\}}
\maketitle
\label{firstpage}

\begin{abstract}
We describe the calculation of the stochastically heated dust emission using the 3D ray-tracing dust radiative transfer code DART-Ray,
which is designed to solve the dust radiative transfer problem for galaxies with arbitrary geometries. 
In order to reduce the time required to derive the non-equilibrium dust emission spectra from each 
volume element within a model, we implemented an adaptive SED library approach, which we tested for the case of axisymmetric galaxy 
geometries. To show the capabilities of the code, we applied DART-Ray to a high-resolution 
N-body+SPH galaxy simulation to predict the appearance of the simulated galaxy at a set of wavelengths from the UV to the sub-mm. 
We analyse the results to determine the effect of dust on the observed radial and vertical profiles of the 
stellar emission as well as on the attenuation and scattering of light from the constituent stellar populations. We also quantify the 
proportion of dust re-radiated stellar light powered by young and old stellar populations, both bolometrically and as a function of infrared 
wavelength.

\end{abstract}


\section{Introduction}

The investigation of the stellar and dust content of galaxies is mainly based on the analysis of the multiwavelength UV-to-submm emission
spectra of the radiation escaping from galaxies and received by ground--based and space observatories. A quantitative understanding of 
the emission detected at these wavelengths, integrated over the entire galaxy or from specific regions within a galaxy, is an essential 
requirement for studies of star formation on galactic scales as well as for studies of galaxy evolution.   
Except for the possible 
presence of a bright active galactic nucleus, the spectral energy distribution (SED) of the UV-to-submm continuum emission of a 
galaxy is determined by the intrinsic distribution of stars and dust within the galaxy: the stars emit 
UV-to-near--IR radiation, whose spectra depends mainly on their intrinsic mass, age and metallicity; the interstellar dust particles absorb 
and scatter a fraction of the radiation coming from stars (or from other dust particles) and re-emit 
the absorbed luminosity in the infrared range. The amount of radiation absorbed and 
scattered by dust particles, as well as their emission spectra, are determined by the relative geometry between stars and 
dust, the optical properties of the dust grains, the dust abundance and the stellar luminosity. 
Therefore, the output radiation SED along a particular line-of-sight within a galaxy is the product of a large variety of physical 
factors.

In principle, at least some of the information contained within the observed emission SED of a galaxy can be extracted by 
comparing observed and predicted SEDs for a set of galaxy models (either analytical or derived from numerical simulations). 
This comparison requires the identification of sufficiently realistic galaxy models and dust radiative transfer (RT) calculations needed to 
predict their emission SEDs. However, this approach is still not commonly adopted for the 
analysis of galaxy data because of the intricacies in either calculating or empirically determining the relative distribution of stars and dust, the large number 
of parameters that need to be constrained and the high computational effort to perform those calculations. Instead, simpler techniques are 
routinely used to derive estimates of intrinsic stellar and/or dust parameters, such as 
combined stellar and dust emission SED fitting (e.g. da Cunha et al. 2008, Devriendt et al. 1999, Marshall et al. 2007, 
Noll et al. 2009, Sajina et al. 2006);  
dust emission SED fitting using theoretically/empirically derived dust emission templates (e.g. Chary \& Elbaz 2001, 
Compi\`{e}gne et al. 2011, 
Dale et al. 2001, Dale \& Helou 2002, 
Draine et al. 2007, Natale et al. 2010) or single temperature modified black body curves (e.g. Amblard et al. 2010, 
Bourne et al. 2012, Dariush et al. 2011,  Smith et al. 2012). In addition, empirically constrained proxies for star formation rates based on a 
variety of observational tracers are in common use (see Buat et al. 2010, Kennicutt \& Evans 2012, Calzetti et al. 2013 and references therein.). 
However, all these techniques are unable to consider 1) the 
 effect of non local contributions to the scattered and dust reradiated starlight at specific positions within a galaxy, 
 2) the intrinsic geometry of stellar and dust distribution and 3) the differences in observed emission due to different galaxy inclinations. 
 All these aspects can be taken into account only by performing RT calculations.   

In order to interpret galaxy SEDs, the use of galaxy models combined with radiation transfer calculations can follow two 
different approaches (see Popescu \& Tuffs 2010, PT10). 
The first approach (called ``decoding observed panchromatic SEDs'' in PT10) consists of 
performing a fit to the observed multiwavelength SEDs of galaxies by using a grid of galaxy models, whose stellar and dust component distributions are described by analytical functions. 
This approach allows the distribution of stars and dust within galaxies to be empirically constrained. 
This method was pioneered and calibrated for nearby edge-on galaxies by Xilouris et al. (1997,1998,1999) in the optical range, and 
by Popescu et al. (2000) for the multiwavelength UV/optical-FIR/submm range. Further applications of this method to individual galaxies have 
also mainly focused on edge-on galaxies, which have been decomposed in terms of axisymmetric stellar 
and dust components, often including a clumpy component and/or spiral arms (see e.g. Baes et al. 2010, Bianchi \& Xilouris 2011, 
Bianchi 2007, de Looze et al. 2012a,b, De Geyter et al. 2013, 2014, MacLachlan et al. 2011, Misiriotis et al. 2001, 
Popescu et al. 2004, Schechtman-Rook et al. 2012, Schechtman--‐Rook \& Bershady 2014). By using this 
approach on a set of edge-on galaxies, those authors derived the large scale parameters of the galaxy stellar and dust distributions, such as 
their scale length/height and the total values for the intrinsic stellar luminosity, dust mass and optical depth. 
This method has been also generalised for application to statistical samples of any orientation and bulge-to-disk ratio (spiral morphology) 
by Tuffs et al. (2004), and Popescu et al. (2011) and has been successfully applied to predict the statistical behaviour of various observables 
in Driver et al. (2007), (2008), (2012), Graham \& Worley (2008), Masters et al. (2010), Gunawardhana et al. (2011),  Kelvin et al. (2012), 
(2014), Grootes et al. (2013, 2014), Pastrav et al. (2013a,b) and Vulcani et al. (2014). 
Recently, De Looze et al. (2014) extended the applicability of the ``decoding technique'' to the modelling of a resolved nearby face-on galaxy. 
Other approaches to the "decoding observed panchromatic SEDs" include the radiative transfer models of Rowan-Robinson \& Crawford (1989),
Siebenmorgen \& Kruegel (1992), Silva et al. (1998), Efstathiou \& Rowan-Robinson (2003), Rowan-Robinson et al. (2005),
Piovan et al. (2006), Efstathiou \& Siebenmorgen (2009), Rowan-Robinson \& Efstathiou (2009), Rowan-Robinson et al. (2010) 
(see also Rowan-Robinson 2012 for a review). 

The second approach (called ``encoding predicted physical quantities'' by PT10) consists of the application of dust radiative transfer codes to 
galaxy models obtained through numerical simulations of galaxy formation and evolution. In contrast to the decoding approach, in this case the 
stellar and dust distributions are derived from first principles. The SEDs of the simulated galaxy have been used by 
several authors to elucidate the dependence of multiwavelength emission from galaxies on their intrinsic physical
properties (e.g. Hayward et al. 2012, 2013, 
Chakrabarti \& Whitney 2009, Jonsson et al. 2010, Dom\'{i}nguez-Tenreiro et al. 2014). 
One advantage of performing radiative transfer calculations for numerically simulated galaxies, which has not been considered very 
much so far, is that one can also study 
the predicted observed properties of these galaxies on local scales. This is possible provided that both the galaxy simulation and the 
RT calculation are performed at sufficiently high spatial resolution. The more detailed distributions
of stars and dust in the numerically simulated galaxies, compared to the simplified galaxy models used in the first approach, makes 
these RT calculations particularly amenable for the study of attenuation and scattering of stellar light as well as of dust heating and 
emission on local scales. However, one should note that the entire approach relies on the input physics
assumed for both the galaxy numerical simulation and the radiation transfer calculations being complete and able to resolve all relevant 
physical processes.  

In this paper we follow the second approach. Specifically, we present an application of DART-Ray, 
a purely ray-tracing 3D dust radiative transfer code presented in Natale et al. (2014), to a high resolution galaxy N-body+SPH simulation.  
This work has the following main purposes. Firstly, we show how we included in DART-Ray an efficient method for the calculation of the 
stochastically heated dust emission by implementing an approach based on the construction of an adaptive library of dust emission SEDs.   
Then, we show the full 3D capabilities of DART-Ray by applying it to predict the emergent radiation from a complex and highly 
resolved model of a galaxy. We use the results of this specific RT calculation to perform an example study of the dust attenuation, 
scattering, heating and emission on both global and local scales, which can be used as a generic framework for future similar studies. 

The structure of the paper is as follows. In section 2 we summarize the main characteristics of the RT algorithm implemented in 
\textsc{DART-Ray} and we present the procedure we implemented to include the calculation of the stochastically 
heated dust emission. In section 3 we describe a method to determine the relative fraction of dust emission powered by young and old stellar 
populations for a given galaxy model (which we use in this work for the analysis of the dust heating properties). In section 4 we describe 
the details of the simulated galaxy RT calculation we perform for this work. In section 5 we present the predicted galaxy maps and we show the 
results for the dust attenuation, scattering and heating properties. 
A summary closes the paper. Note that additional plots and figures for several sections can be found in the appendix.

\section{Calculation of dust emission using \textsc{DART-Ray}}
The calculation of dust emission requires an accurate knowledge of the radiation fields through the entire volume of the galaxies, including regions 
of low illumination. Indeed, far-infrared(FIR)/submm observations of spiral galaxies show that most of the dust emission luminosity
is emitted longwards of 100 micron (see e.g. Sodroski et al. 1997, Odenwald et al. 1998, Popescu et al. 2002, 
Popescu \& Tuffs 2002, Dale et al. 2007, 2012, Bendo et al. 2012) through grains situated in the diffuse ISM, which are 
generally located at very considerable distances from the stars heating the dust. In addition, infrared radiation shortwards of 100\,$\mu$m is 
also known to have a significant contribution from small grains out of equilibrium with the radiation
fields of low energy densities, the so--called stochastically heated grains. One of the motivations for developing \textsc{DART-Ray} was 
specifically to provide a tool for accurately determining radiation fields in galaxies for the calculation of dust emission. 
The determination of the radiation fields has been described in Natale et al. (2014), 
together with the detailed description of the RT algorithm. 
In this section we briefly recap some of the main features of the RT algorithm (Sect.\ref{DR_RT_algo}), before describing the dust emission calculations
including the stochastically heating mechanism (Sect.\ref{non_LTE_dust_sect}).

\subsection{The DART-Ray 3D dust RT algorithm}
\label{DR_RT_algo}

\textsc{DART-Ray} is a purely ray-tracing 3D dust 
radiative transfer code, described in detail in Natale et al. (2014)
\footnote{In particular, see section 2 and 3 of Natale et al. (2014) for a brief description of the RT algorithm and for the complete technical details 
of its implementation.}. The algorithm implemented in this code shares with the more common 
3D Monte-Carlo RT algorithms the aim of determining the contribution of each radiation source to the
radiation field energy density (RFED) in regions within the model volume where this contribution is ``significant'' 
(in the sense that 
it cannot be neglected in order to calculate accurately the RFED at a certain position). 
However, instead of using random numbers and probability functions to determine the location of the emission and the propagation of photons 
within the model volume, the RT algorithm implemented in \textsc{DART-Ray} exploits the concept of ``source influence volume'' to decide where 
to follow the propagation of radiation from each source. The ``source influence volume'' is defined 
as the volume which is significantly illuminated by a source of direct\footnote{By ``direct radiation'' or ``direct light''
we mean the radiation emitted by stars in a certain direction which is neither absorbed or scattered by dust. Thus, direct radiation 
reaches the observer or a certain position within the model volume while propagating along the same initial direction.} or scattered radiation 
within a galaxy model. 
The \textsc{DART-Ray} algorithm   
identifies regions including the influence volume for each source and calculates the source RFED contributions only within them. 
This procedure becomes gradually more efficient when sources of higher order scattered light are 
considered, since their influence volumes tend to become smaller and smaller at each scattering iteration. On the other hand, for the 
purpose of calculating the 
escaping radiation from the model, the code calculates for each volume element (``cell'') how much of the initial 
monochromatic specific intensity $I_{\lambda,0}$, determined by direct or scattered radiation sources within the volume element, 
is escaping from the system.  
This is done by multiplying $I_{\lambda,0}$ by a factor $e^{-\tau_\lambda(\theta,\phi)}$, where $\tau_\lambda(\theta,\phi)$ is the 
optical depth from the cell to the border of the model along the direction ($\theta,\phi$) (this procedure 
is similar to the so called ``peel-off'' technique implemented in modern Monte-Carlo codes to make them more efficient in creating galaxy images, 
see review by Steinacker et al. 2013).
The specific intensity of the radiation escaping from the model is calculated for a set of directions, specified by the user in the input, 
and is stored for each cell. This can then be used to create galaxy images by using 3D volume rendering. Further characteristics of the 
\textsc{DART-Ray} code include an 
adaptive Cartesian 3D spatial grid, an iterative procedure to optimize the angular density of the rays launched from 
each cell, the treatment of anisotropic scattering and parallelization using OpenMP. 

Each run of the code is performed separately for each wavelength.
In the output of the code both the monochromatic RFED and the outgoing radiation specific intensity for each cell are 
provided for the direct and scattered radiation separately. The total RFED distribution can then be used to calculate the dust emission from 
each cell, assuming that the dust is in thermal equilibrium with the radiation field or is stochastically heated. In the next section, 
we describe the method implemented to perform the dust emission calculation in the latter case.

\subsection{Inclusion of non-equilibrium dust emission calculations in DART-Ray}
\label{non_LTE_dust_sect}
In order to obtain realistic predictions of the dust emission SED in galaxies, an important physical phenomena that has to be taken 
into account is   
the stochastic heating of the dust grains which regards dust  
particles out of equilibrium with the heating radiation field. In this case, the emission SED from each grain cannot be 
modelled as a single 
temperature modified black body spectrum. Stochastical heating cannot be neglected when the energies carried by the photons absorbed 
by the grain are comparable to, or exceed, the grain enthalpy. This occurs when the grain cooling times are shorter than the 
time needed for another photon to be absorbed. These conditions can be realized when grains have very small 
sizes (that is, $10-100$ \AA{}), the radiation fields are weak (as in the case of the local interstellar radiation field) or because of the 
hardness of the radiation field spectra (see Siebermorgen et al. 1992). As a result, dust grains can experience large temperature 
fluctuations and their infrared emission tends in general to be warmer than expected from their equilibrium 
temperature. 

The calculation of the emission spectra from stochastically heated grains consists of two steps. In the first step, one has to 
determine the probability distribution for the temperature of grains of different sizes and composition. In the second step, one calculates the 
emission from each grain based on the inferred temperature probability distribution and integrates over the assumed dust size 
distribution and the different grain species. In this work we perform these calculations as in Popescu et al. (2011, hereafter P11).
The temperature probability distribution is calculated following Voit (1991), based on the numerical integration method
of Guhathakurta \& Draine (1989). The calculation for stochastically heated grains is performed for all grain sizes, except for 
grains with very narrow temperature probability distribution which are assumed to be at equilibrium. A Gaussian approximation for the 
temperature distribution of grains close to equilibrium (see Voit 1991) is used to determine the transition between equilibrium and 
stochastical heating regimes. 
We also assumed the dust model 
of Weingartner \& Draine (2001) for the Milky Way dust (with $R_V=3.1$), modified by Draine \& Li (2007). 

For a given dust model and radiation field energy density spectra, the calculation of the stochastically heated dust emission 
requires of order of few seconds using a single processor on a typical modern computer. This calculation time is not 
a concern when the calculation is performed for 2D galaxy models, where the 
number of grid points is of order of $10^3-10^4$. However, a brute force approach, where the exact calculation is performed at each 
grid point, becomes inconvenient for 3D galaxy models, where the number of grid points can be of order of $10^5-10^6$. 

Various authors have followed different strategies to reduce the time needed for dust stochastic heating calculations, 
once a radiation field distribution has been inferred using a radiative transfer code. Siebenmorgen et al. (1992) and 
Misselt et al. (2001) assume that stochastic heating 
needs to be calculated only for grains smaller than a certain size ($\approx 80-100$ \AA{}). Bianchi et al. (2008) adopt an approach 
using the same temperature bins for all grid points and grain sizes, when computing the temperature probability distribution. 
This allows the tabulation of some quantities often used in the calculation.  
Wood et al. (2008) and Jonsson et al. (2010) use libraries of precomputed dust emission SEDs and assume that the dust emission is 
determined by the amplitude but not by the spectral shape of the heating radiation. Finally, Baes et al. (2011), 
further developing an approach first proposed by Juvela \& Padoan (2003), construct a dust emission SED adaptive library for a given 
RT calculation, which is derived by grouping the inferred 
radiation field spectra in bins of similar equilibrium dust temperatures and effective wavelengths (see their section 3.5.2). 
Then they calculate the dust emission spectra from the average of the radiation field spectra in each bin. 
The use of these two parameters to construct the SED adaptive library allowed those authors to approximately consider both the intensity and 
spectral shape of the radiation field, unlike the simpler approach of Wood et al. (2008) and Jonsson et al. (2010). 

As in Baes et al. (2011), we use a dust emission SED adaptive library approach which can be built for a given radiation field 
energy density distribution $U_\lambda$ inferred by a radiative transfer calculation performed on a certain galaxy model. However, 
in our approach we consider as the parameters 
to be binned the wavelength--integrated radiation field energy densities in the UV and in the optical/infrared regime. These are\footnote{Note 
that for the moment we consider only the radiation field produced by stellar light in the evaluation of $U_{UV}$ and $U_{opt}$. 
This is because we are neglecting dust self-heating in our calculations. This should be a good approximation for galaxies, since they are 
typically optically thin in the infrared. However, if this effect were to be included, 
our method could be extended to include the integrated infrared radiation field energy density as a separate parameter.}:
\begin{equation}
 U_{\rm UV}=\int_{918 \AA{}}^{4430 \AA{}}U_\lambda d\lambda
\end{equation}
and
\begin{equation}
 U_{\rm opt}=\int_{4430 \AA{}}^{5\mu m}U_\lambda d\lambda.
\end{equation}
Unlike Baes et al. (2011), this approach explicitly takes into account the optical-to-UV ratio of the radiation field intensity at each 
position. The motivation for our approach is that for any fixed equilibrium dust
temperature there can be a range of ratios of UV to optical light, each yielding a different dust emission SED.  

Our method consists of the following steps:\\
1) we calculate the total UV and optical/infrared radiation field energy densities, $U_{\rm UV}$ and $U_{\rm opt}$ at each grid volume 
element (``cell''); \\
2) we bin the inferred range of values covered by $U_{\rm UV}$ and $U_{\rm opt}$;  \\
3) we consider all the grid cells whose $U_{\rm UV}$ and $U_{\rm opt}$ values fall into the same two-dimensional bin 
and we average their radiation field spectra; \\
4) we calculate for each $\left[U_{\rm UV},U_{\rm opt}\right]$ bin the stochastically heated dust emission spectra per unit dust mass by using 
the average radiation field spectra calculated in step 3;  \\
5) we assign to each cell the dust emission spectra, scaled by the cell dust mass, derived for the $\left[U_{\rm UV},U_{\rm opt}\right]$ bin 
corresponding to the cell $U_{\rm UV}$ and $U_{\rm opt}$ values.  

In order to test the accuracy of this approximation, we have compared the results of the time-consuming brute 
force approach (where the calculation is performed exactly for the RFED spectra of each cell) with those obtained using the SED 
adaptive library approach described above. Specifically, we tested these two approaches for two different RFED distributions,
derived assuming the axisymmetric galaxy model geometry defined in P11. In the first test, we considered the RFED distribution calculated by including only the 
thick stellar disc (we set the luminosity scaling parameters ``old'' and ``SFR'' to old=1 and SFR=0\,$M_\odot/yr$, see P11 for definitions). 
In the second test, we considered the RFED distribution derived by including only the thin stellar disc 
(parameters old=0 and SFR=1\,$M_\odot/yr$). 
For both tests we assumed that the dust is distributed within two dust discs (the thick and the thin
dust disc from P11) with a total face-on central optical depth in the B-band
$\tau_B^f=1$. We derived the RFED distributions for the two tests by performing radiative transfer calculations at all wavelengths 
tabulated in Table E.2 of P11. Then, we calculated the stochastically heated dust emission using both the brute force and the 
adaptive library approach for each of the two cases.  

\begin{figure}
\centering
\includegraphics[scale=0.40]{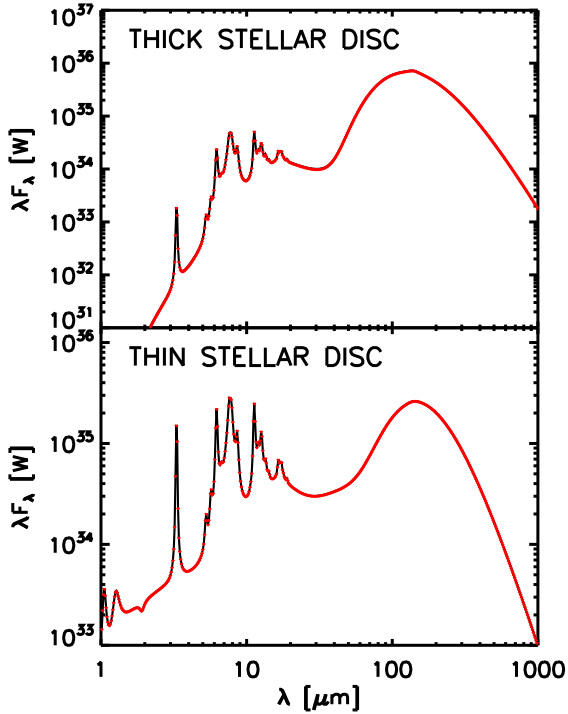}
\caption{Comparison of the integrated dust emission SEDs calculated using the brute force approach (black line) and the adaptive library approach
(red points). The top panel shows the SEDs for the test including only the thick stellar disc while the bottom one those for the 
test including only the thin stellar disc (see text for details). Both the brute force approach and the library approach adopt the 
same logarithmic sampling of the SED through the whole spectral range, with the sampling interval $\Delta\log\lambda=0.0059$.} 
\label{int_sed_comp}
\end{figure}

\begin{figure}
\centering
\includegraphics[scale=0.40]{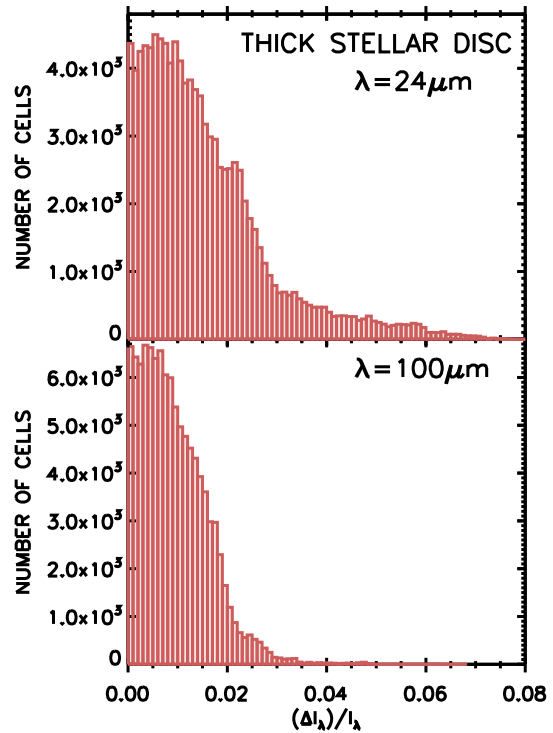}
\caption{Histograms showing the number of cells as a function of the relative difference of the 24 and 100 $\mu$m fluxes calculated 
using the brute force and the adaptive library approach for the thick stellar disc test.} 
\label{rel_diff_old}
\end{figure}

\begin{figure}
\centering
\includegraphics[scale=0.40]{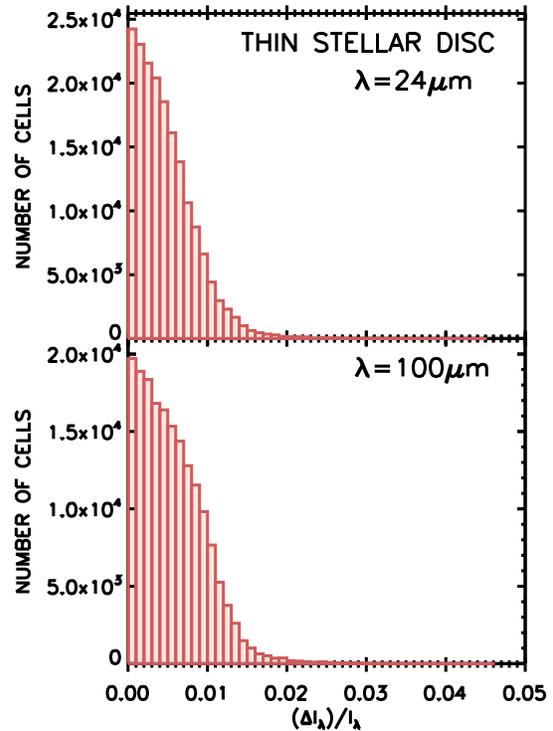}
\caption{Same as Fig.\ref{rel_diff_old} but for the thin disc test.}  
\label{rel_diff_sfr}
\end{figure}

Since the thick stellar disc does not emit in the UV (see table E.2 of P11), in the test including only this component we considered 
 only $U_{opt}$ when constructing the dust emission library. The top panel of Fig.\ref{int_sed_comp} shows the integrated dust emission SEDs for 
both the brute force approach and the SED adaptive library approach for this test. The integrated SEDs are remarkably close to each other. 
In Fig.\ref{rel_diff_old} we also show for this test, where the illumination is only by the thick stellar disc,
histograms of the number of cells with a given relative difference between the results of the two methods 
for the dust emission flux at 24 and 100\,$\mu m$. This shows that there are differences of order of a few percent for the 
predicted fluxes for the single cells. 
In the test including only the thin stellar disc, we calculate both $U_{UV}$ and $U_{opt}$
since the thin stellar disc emits from the UV to optical/infrared wavelengths. The bottom panel of Fig.\ref{int_sed_comp} shows the 
integrated SED obtained by the two methods for this test and Fig.\ref{rel_diff_sfr} shows the corresponding histograms of the 
number of cells as a function of the 24 and 100$\mu$m relative flux differences. 
The accuracy achieved is comparable with 
that of the previous test. For both the thick and thin stellar disk we also derived histograms of the numbers of cells as a function of the 
8 $\mu$m relative flux differences (not shown in this paper), which we find to be similar to the histograms at 24 micron. 
The results of these tests suggest that the adaptive library approach we designed is a reliable shortcut to 
calculate the stochastically heated dust emission for galaxy models. In fact, this method allows us to calculate a factor 10-100 
less dust emission spectra than by using the brute force approach but still preserving a good accuracy for the final results.   

However, one should also note that the use of 
axisymmetric distributions of stellar emissivity and dust opacity for these tests simplifies the distribution of RFED within the 3D 
galaxy model, since 
there will be many cells having exactly the same radiation field spectra. Nonetheless, this test suggests that it is quite accurate
to use averages of similar RFED spectra instead of using the exact 
values for each cell. In this case only a few percent difference for the monochromatic dust emission is found on a cell--by--cell basis 
while the integrated SED is recovered to within 1\% accuracy. In addition, within a real galaxy a certain degree of axisymmetry and/or 
similarity in the distribution of the RFED spectra is expected.   

\section{Calculation of the dust emission fraction powered by young and old stellar populations}
\label{sect_calc_y_frac}

It is of great interest to know which stellar populations (SPs) are responsible for powering the dust emission in galaxies. Many authors have
tried to address this question, mainly relying on simple geometrical comparisons between the surface brightness distributions at different
wavelengths (e.g. Boquien et al. 2011, Bendo et al. 2012). Such comparisons make the implicit assumption that the stellar emissivity within 
the resolution element of the telescope dominates the heating of the dust which is visible in the same telescope beam. However, this 
assumption will break down when radiation fields are not dominated by local sources, but instead by distributed sources, offset by 
more than the resolution element. Quantitative determination of radiation fields provided by all 
contributing sources can only be achieved by following the transfer of radiation within these objects. In this context \textsc{DART-Ray} 
can be used as a generic tool to examine exactly this question.

In this section we describe a general method to evaluate which stellar populations are responsible for powering the dust emission 
as a function of position and infrared wavelength. This method is in principle applicable to any radiation transfer code and 
any galaxy model/simulation. The method we present here is an extension of the procedure used by Popescu et al. (2000) to elucidate the 
fraction of the dust emission in NGC891 powered by UV photons as a function of infrared wavelength. The procedure we adopt here involves the 
calculation of the radiation fields due to all the stellar populations in the galaxy, but also additional calculations of the radiation 
fields due to either young or old SPs.

The input grid for the additional RT run can be created by considering only the emission from stellar populations with ages younger or older 
than a threshold age $t_{\rm lim}$.   
In this way one can determine the RFED contributions $U_{\rm \lambda,young}^{\rm cell}$ and $U_{\rm \lambda,old}^{\rm cell}$ due to young and 
old SPs to the total RFED in each cell (one contribution can be derived directly from the new RT run and the other by subtraction to the 
total RFED). 
From the values of $U_{\rm \lambda,young}^{\rm cell}$ and $U_{\rm \lambda,old}^{cell}$ one can then derive the luminosity coming from young and old SPs which 
is absorbed in each cell:
\begin{equation}
 L^{\rm cell,abs}_{\rm young}=cM_{\rm d}^{\rm cell}\int {k_{\rm \lambda,abs}U_{\rm \lambda,young}^{\rm cell}d\lambda}
\end{equation}
and 
\begin{equation}
 L^{\rm cell,abs}_{\rm old}=cM_{\rm d}^{\rm cell}\int {k_{\rm \lambda,abs}U_{\rm \lambda,old}^{\rm cell}d\lambda}
\end{equation}
where $k_{\rm \lambda,abs}$ is the absorption coefficient per unit dust mass for the adopted dust model and $M_{\rm d}^{\rm cell}$ is the 
dust mass in the cell. Instead, the total absorbed luminosity in each cell $L^{\rm cell,abs}_{\rm TOT}$ is equal to the sum of the above 
absorbed luminosities:
\begin{equation}
 L^{\rm cell,abs}_{\rm TOT}=L^{\rm cell,abs}_{\rm young}+L^{\rm cell,abs}_{\rm old}
\end{equation}
Since dust simply re-emits all the luminosity that it absorbs, the total bolometric dust emission luminosity and the contributions powered 
by young and old SPs are equal to $L^{\rm cell,abs}_{\rm TOT}$, 
$ L^{\rm cell,abs}_{\rm young}$ and  $L^{\rm cell,abs}_{\rm old}$ respectively. Therefore, one can use those quantities to calculate the relative 
fractions of dust emission luminosity powered by the two SPs for each cell as well as for the entire model. 

While determining the amount of the dust bolometric luminosity powered by young and old SPs is quite straightforward, it is rather more complex 
to determine their relative contributions to the dust emission luminosity at single infrared wavelengths. In contrast to 
the bolometric case, one cannot simply calculate the amount of dust emission by considering $U_{\rm \lambda,young}^{\rm cell}$ and $U_{\rm \lambda,old}^{\rm cell}$ 
separately and then sum the inferred dust  
emission spectra in order to obtain the total dust emission. This is because the dust emission spectra of each grain depends non linearly from the 
RFEDs and, thus, it has to be calculated from the total RFED (given by the sum of $U_{\rm \lambda,young}^{\rm cell}$ and $U_{\rm \lambda,old}^{\rm cell}$). 
Also, when calculating the relative fraction of monochromatic luminosity powered by young and old SPs, it is necessary to consider the 
emission spectra from the single grains. The reason is that young and old SPs emit predominantly in different regions of the UV to near-IR wavelength range and, at 
the same time, dust grains have wavelength dependent absorption coefficients determined by their size and composition. Therefore, different 
grains can be more or less efficient in absorbing radiation from young or old SPs. Finally, when considering the emission spectra of a single grain, 
one can assume that the 
fraction of the emission at a single wavelength powered by young/old SPs is equal to their relative contribution to the single grain bolometric
emission. This is because one can assume that all photons heat the grain simultaneously and the absorbed luminosity is 
simply redistributed in the grain emission spectra\footnote{Strictly speaking this statement is true only for grains in equilibrium with
the UV/optical radiation fields. For stochastically heated grains this is only an approximation, which however gives rise to only minor 
deviations from the predictions of our present analysis.}. For all these reasons, one can use the following procedure for each single cell: \\

1) Calculate the monochromatic luminosity $L_\lambda^{\rm cell}(a,i)$ at wavelength $\lambda$ as well as the bolometric luminosity $L^{\rm cell}(a,i)$ for each grain of 
size $a$ and grain species $i$, assuming the total RFED value (including contributions from both young and old SPs);

2) Calculate the bolometric dust luminosity powered by young stellar 
populations $L_{\rm young}^{\rm cell}(a,i)$ for each grain size and species. This can be calculated from the heating rate $H_{\rm young}^{\rm cell}(a,i)$ 
derived from the RFED from the young SP run:
\begin{equation}
 L_{\rm young}^{\rm cell}(a,i)=H_{\rm young}^{\rm cell}(a,i)
\end{equation}

3) Calculate the fraction of bolometric single grain luminosity powered by young SPs $\chi_{\rm young}(a,i)$, which is given by:
\begin{equation}
 \chi_{\rm young}^{\rm cell}(a,i)=\frac{L_{\rm young}^{\rm cell}(a,i)}{L^{\rm cell}(a,i)}
\end{equation}

4) Assuming that $\chi_{\rm young}^{\rm cell}(a,i)$ is equal to the fraction of heating by young SPs at all wavelengths of the single grain emission 
spectra, one can then derive the single grain monochromatic emission luminosity powered by young SPs $L_{\lambda,\rm young}^{\rm cell}(a,i)$ as: 
\begin{equation}
 L_{\lambda,\rm young}^{\rm cell}(a,i)=\chi_{\rm young}^{\rm cell}(a,i)L_\lambda^{\rm cell}(a,i)
\end{equation}

5) By integrating $L_{\rm \lambda, young}^{\rm cell}$ over the grain size distribution and summing over the different species, one obtains the total amount 
of monochromatic dust luminosity powered by young SPs $\Gamma_{\rm \lambda, young}^{\rm cell}$:
\begin{equation}
 \Gamma_{\rm \lambda, young}^{\rm cell}=\sum_i\int {n(a,i)L_{\rm \lambda, young}^{\rm cell}(a,i)da}
\end{equation}
where $n(a,i)$ is the grain number density. By combining $\Gamma_{\rm \lambda, young}^{\rm cell}$ with the total value for the monochromatic dust emission 
luminosity, one can derive the fraction of monochromatic dust emission powered by young and old SPs.

\section{Application to a high resolution N-body+SPH galaxy simulation}

In this section we describe the galaxy simulation used in this paper and provide details on how the 3D grids of stellar emissivity and dust 
opacity, input to the RT calculations, have been created. We also describe the methods we used to analyse the simulated galaxy maps, as well as 
the maps of the intrinsic distributions of stellar emission, dust opacity and gas mass. 

\subsection{The N-body+SPH galaxy simulation}
\label{det_vgal}

The N-body+SPH galaxy simulation we used is described in detail by Cole et al. (2014) and references therein. Here we summarize only 
a few of the main details of particular interest for this work. The galaxy simulation
 models the formation and 
evolution of a disc galaxy within a corona of hot gas embedded in a dark matter halo. 
The simulation was run using the N-body+SPH code \textsc{GASOLINE}.  Subgrid prescriptions were used for the treatment of 
 warm/hot gas cooling, star formation and stellar feedback (following Stinson et al. 2006, see Cole et al. 2014). 
Metal cooling has not been considered in this calculation. 
Metal enrichment of the gas particles is calculated by assuming the yields of Woosley \& Weaver (1995).
The adopted force softening 
parameter, setting the spatial resolution of the calculation, is 50\,pc. We note that no stellar particle is present 
at the beginning of the simulation but all stars are born out of cooled gas whose density is high enough to trigger star formation.
For the RT calculation, we considered the model after 10 Gyr of evolution. At this epoch, Cole et al. (2014) show that the model has a 
prominent nuclear disc while Ness et al. (2014) show that the bulge has a range of stellar ages. The galaxy global parameters can be found in 
Table \ref{table_galaxy param}. The total stellar luminosity and dust mass are derived as described in the following section 
\ref{input3d_sect}.  

\begin{table}
\begin{center}
\caption{Simulated galaxy parameters.}
\begin{tabular}{l | l}
\hline\hline
Parameter & Value \\
\hline
$M_{\rm stars}$ & $6.43\times10^{10}$\,M$\odot$ \\
$M_{\rm gas}$ & $4.12\times10^{9}$\,M$\odot$ \\
$M_{\rm dust}$ & $1.14\times10^{7}$\,M$\odot$ \\
$L_{\rm stars,TOT}$ & $2.57\times10^{44}$ erg/s \\
\hline
\end{tabular}
\label{table_galaxy param}
\end{center}
\end{table}

\subsection{Creation of the input 3D adaptive Cartesian grid} 
\label{input3d_sect}

In order to perform the RT calculation with \textsc{DART-Ray}, the first step consists 
of the creation of a 3D adaptive Cartesian grid, containing the information on the stellar volume emissivity and dust opacity distribution of the galaxy
model. This grid has been constructed from the output of the N-body+SPH galaxy simulation, which includes the position coordinates, mass and metallicity 
for both gas and stellar particles of the simulated galaxy after $t=10$\,Gyr of evolution. In addition, the galaxy simulation output provides the ages 
of the stellar particles and the temperature for the gas particles.  

For each of the cells used to subdivide the volume considered in the RT calculation, one needs to derive the extinction coefficient $\kappa_\lambda$
 and the stellar volume emissivity $j_\lambda$ at several wavelengths from UV to optical/near-infrared. 
The extinction coefficient $\kappa_\lambda$ is equal to the product $C_{\lambda}^{\rm gas}\rho_{\rm gas}$, where 
$C_\lambda^{\rm gas}$ is the dust cross section per unit gas mass\footnote{$C_\lambda^{\rm gas}=C_\lambda/(1.4\times m_H)$ where 
$C_\lambda$ is the dust cross section per unit Hydrogen atom, see Appendix A of Popescu et al. 2011, and $m_H$ the hydrogen mass.} and 
$\rho_{\rm gas}$ is the gas mass density. One needs to know this product for each cell in order to 
calculate the decrease of the 
specific intensity $I_\lambda$ along a ray crossing the cell. In fact, the new value of $I_\lambda$ after a cell 
crossing is given by:
\begin{equation}
 I_{\lambda,i+1}=I_{\lambda,i}e^{-C_\lambda^{\rm gas}\rho_{\rm gas} l_c}
\end{equation}
where $l_c$ is the ray crossing path within a cell\footnote{Note that the increase of $I_\lambda$ due to the volume emissivity $j_\lambda$ along
the crossing path is not considered in this formula. In the \textsc{DART-Ray} algorithm, when following the variation of $I_\lambda$ along a 
certain ray direction, the volume emissivity is considered only for the cell originating the ray (see section 3.2 of Natale et al. 2014).}.  

For a given cell, we calculated $\rho_{\rm gas}$ in the following way. Firstly, we considered all the gas particles located 
within the cell with temperatures less then 10$^6$ K (we assumed gas particles with higher temperatures are dust-free, since dust in hot gas is 
destroyed by sputtering within short time scales, see e.g. Draine \& Salpeter 1979). Then, we divided the total mass of the selected gas particles 
by the cell volume in order to obtain the average gas mass density in the cell. 

To calculate the value of $C_\lambda^{\rm gas}$ for each cell, we considered the extinction efficiencies $Q_\lambda$ and the size 
distribution from the dust model of Weingartner \& Draine 2001 (WD01),  modified by Draine \& Li 2007 (DL07), consistently with the dust 
emission calculations we perform once the RFED has been derived (see section \ref{non_LTE_dust_sect}). 
However, we also took into account the variation of the gas metallicity throughout the galaxy model. In fact, this parameter is one of the 
factors affecting the dust-to-gas mass ratio and, thus, the dust cross sections per unit gas mass. From the N-body+SPH simulation we do not 
obtain a complete tracing of the abundance for each chemical element. We considered only the iron abundance relative to solar abundance, 
that is, $[Fe/H]=\log(Fe/H)-\log(Fe/H)_\odot$. 
 We assumed that [Fe/H] traces the metallicity Z according to $Z=Z_{\rm MW}10^{[Fe/H]}$ where $Z_{\rm MW}$ is the metal abundance of the Milky Way,
 assumed to be equal to 0.018 (see footnote\footnote{The specific value of $Z_{\rm MW}$ is unimportant for the calculation of $C_\lambda^{\rm gas}$, since 
 it cancels out in equation \ref{c_lambda}, but it is used later for determining the absolute metallicity of the stellar particles.}).  
 Then, we assumed that the dust cross section $C_\lambda^{\rm gas}$ is 
proportional to the metal abundance through a linear relation\footnote{The relation between dust-to-gas mass ratio and metallicity is 
linear for massive and high-metallicity galaxies, and becomes steeper for low-mass low-metallicity galaxies (Remy-Ruyer et al. 2014). 
Since our simulated galaxy is massive, we adopt a linear relation between dust-to-gas mass ratio and metallicity.}, 
that is:
\begin{equation}
 C_{\lambda}^{\rm gas}=C_{\lambda,MW}^{gas}*\frac{Z}{Z_{MW}}
 \label{c_lambda}
\end{equation}
where $C_{\lambda,MW}^{\rm gas}$ is the dust cross section per unit gas mass derived for the WD01/DL07 dust model\footnote{For 
diagnostic purposes
we also derived the dust-to-gas mass ratio $\left(M_{\rm dust}/M_{\rm gas}\right)$ for each cell using an expression analogous to formula 
\ref{c_lambda} and 
assuming a Milky Way value $\left(M_{\rm dust}/M_{\rm gas}\right)_{\rm MW}=0.008$:
$ \left(M_{\rm dust}/M_{\rm gas}\right)=\left(M_{\rm dust}/M_{\rm gas}\right)_{\rm MW}*\frac{Z}{Z_{MW}}$. This parameter has been used to 
derive the total dust mass shown in Table \ref{table_galaxy param}. We note that the assumed values for the Milky Way metallicity 
and dust-to-gas mass ratio imply that the fraction of metals locked into dust grains is 0.44.}. 
Since the metallicity can be different even for gas particles close to each other, in order to calculate the $C_\lambda^{\rm gas}$ value for 
a cell embedding many gas particles, we calculated the mass-weighted average value $\left<Z\right>$ for the metallicity of the gas particles within the 
cell and used equation \ref{c_lambda} with $Z=\left<Z\right>$. We note that from the WD01/DL07 dust model we also obtained the albedo $\omega_\lambda$ 
and the Henyey-Greenstein phase function parameter $g_\lambda$, which are used to calculate fractions of absorbed and scattered light as well 
as the angular distribution of the scattered light intensity after a ray has crossed a cell (see section 3.2 of Natale et al. 2014). 

For each cell one also needs to calculate the average volume emissivity $j_\lambda$ of the stellar radiation emitted within it. 
This quantity is needed to calculate the initial specific intensity $I_{\lambda,0}$ of the ray originated by each cell containing stellar particles
(that is, the value at the position of the ray intersection with one of the faces of the first cell). 
For this purpose, we considered all the stellar particles within a given cell and we multiplied their masses by the luminosity-to-mass ratios 
at different wavelengths obtained from the standard output of the spectral synthesis code Starburst99 (assuming an instantaneous 
burst with Initial Mass Function from 
Kroupa 2001 and the Padova AGB tracks). To calculate the luminosity-to-mass ratios, we assumed that the stellar populations 
within each star particle can be modelled as a single age stellar population with age equal to the stellar particle age and 
metallicity determined by the stellar particle Iron abundance, that is,  $Z=Z_{\rm MW}10^{[Fe/H]}$ as before for the gas particles. 

The cell grid has been created iteratively by a series of subdivisions of the cubic parent cell, embedding the entire volume considered in the RT 
simulation (linear size $\approx$19\,kpc), in child cells of progressively smaller sizes. Each subdivision leads to the creation of 
3$\times$3$\times$3 new child cells. In order to create the grid, several criteria need to be chosen to decide whether a cell 
has to be subdivided or not during the grid creation process. Although one can create different grids for each of the wavelengths used 
in the RT calculation, for simplicity we used subdivision criteria only to create the B-band (4430 \AA{}) grid. Successively, we calculated 
for each cell derived for the B-band grid the values of $k_\lambda\rho_{\rm gas}$ and $j_\lambda$ for all the other wavelengths. 
The chosen subdivision criteria for the B-band are the following:\\ 
(i) minimum subdivision level equal to 4 (cell size = 234\,pc) \\
(ii) maximum subdivision level equal to 6 (cell size = 26\,pc, comparable to the spatial resolution of the N-body+SPH calculation) \\
(iii) maximum cell optical depth $\tau_{B}$= 0.02 \\
(iv) maximum cell stellar luminosity equal to $10^{-4}$ times the galaxy total stellar luminosity. \\
The final grid resulted in about 1.3 million cells. Note that the criteria (iii) and (iv) are not guaranteed to be fulfilled by each grid cell, 
in case they are in conflict with condition (ii). We found that 2.5\% of the cells in the 3D grid are optically thick ($\tau_{B}> 1$). 
It is inevitable to have at least some optically thick cells in the grid, since these correspond to regions of high gas density 
(that is, cold clouds in the ISM). We calculated the grid and performed the RT calculation for 15 wavelengths spanning the range from 912 
\AA{} to 5$\mu$m (as in table E.2 of P11). For each wavelength, we calculated the outgoing radiation specific intensity for a set of directions corresponding to 
the phase angle $\phi=0$ and inclination $\theta$ equal to eight values linearly spaced in the range $[0,\pi/2]$ (from face-on to edge-on).

\subsection{Derivation of radial and vertical profiles from the galaxy maps}
\label{sect_def_prof}

After performing the RT calculation, we were able to create galaxy maps for the stellar and dust emission using 3D volume rendering. This 
technique consists of projecting the emission originating from each cell and escaping outside the model upon planes at different 
inclinations. The same technique can be also used to create 
maps of other quantities such as the projected dust optical depth or the gas mass surface density.
All maps were created by using pixels whose size corresponds to $\sim$3\,pc in physical units. 

In order to analyze the distribution of intensity on each map, we calculated average, density and cumulative profiles as a 
function of galaxy radial distance $R$ or vertical height $z$. These profiles were derived by appropriately averaging or summing the pixel 
values on face-on, inclined and edge-on galaxy maps. Since their derivation is 
independent of the specific quantity considered, we describe here for reference the 
methods we used to obtain those profiles and specify the terminology which will be used throughout the result section.  

For the face-on maps, radial profiles have been obtained by subdividing the galaxy into circular annuli of 
fixed radial width ($\Delta R\sim$100\,pc) with origin at the galaxy centre, and by considering the pixels values in each annulus. 
We calculated the following three types of radial profile: \\
(i) ``average radial profile'': calculated by averaging the pixel values $I_j$ in each i$^{th}$ annulus:
\begin{equation}
 f(R_i)=\frac{\sum_{j=1}^{n(R_i)}{I_j}}{n(R_i)}
\end{equation}
where the sum is performed over all the pixels $j$ such that their radial distance is in the range ($R_i,R_i+\Delta R$) and $n(R_i)$ is the 
number of pixels in the i$^{th}$ annulus. We use this kind of profile to determine how the average intensity of a quantity varies with radial 
distance;\\
(ii) ``radial density profile'': calculated by summing the pixel values $I_j$, multiplied by the pixel physical area $A_p$,
in each i$^{th}$ annulus and then dividing by the annulus radial width $\Delta R$, that is:
\begin{equation}
 f(R_i)=\frac{A_p\sum_{j=1}^{n(R_i)}{I_j}}{\Delta R}
\end{equation}
In this case, $f(R)$ is a distribution function describing how the total intensity on the map is 
distributed radially. That is, it is such that $f(R)dR$ is equal to the total intensity within a ring of inner radius $R$ and 
radial width $dR$;\\
(iii) ``cumulative fraction radial profile'': calculated by summing the pixel values for all pixels at radial distance $r<R_i+\Delta R$ and then 
dividing by the total intensity on the map $I_{\rm TOT}$, that is:
\begin{equation}
 f(R_i)=\frac{\sum_{j=1}^{N(R_i)}{I_j}}{I_{\rm TOT}}
\end{equation}
In contrast to the first two types of profile, in this case $N(R_i)$ is the total number of pixels in the circular area limited by 
the outer boundary of the i$^{th}$ annulus. We used this profile to show the radial ranges where most of the spatially integrated intensity is embedded.\\

For the inclined maps, we calculated the same radial profiles as for the face-on maps with the difference that we used elliptical annuli to subdivide
the galaxy. The ellipses represent points in the galaxy having the same physical radial distance. Their minor--to--major axis ratio $b/a$ is 
determined by the galaxy inclination $\Theta$: $cos{\Theta}= b/a$. 

Finally, for the edge-on maps we calculated profiles as a function of the distance from the galactic plane $z$ by subdividing the map in a series 
of rectangular layers parallel to the galactic plane and whose sizes increase logarithmically with $z$ ($\Delta \log{z}=0.05$\,dex). By considering the pixels in 
each layer, we derived the following profiles: \\
- ``average vertical profile'': calculated by averaging the values of the pixels in the layers at the same distance from the galaxy plane, 
that is:
\begin{equation}
 f(z_i)=\frac{\sum_{j=1}^{n(z_i)}}{I_j}{n(z_i)}
\end{equation}
where the sum is performed over all pixels $j$ within the range of vertical distances $(z_i,z_i+\Delta z)$ and $n(z_i)$ 
is the total number of pixels considered in the average. Note that two layers of pixels are considered in each average, the one above and the one below the galaxy 
plane;\\
- ``vertical density profile'': derived by summing the pixel values, multiplied by the pixel physical area $A_p$, in two 
layers equidistant from the galaxy plane and dividing by the 
layer height $\Delta z_i$, that is: 
\begin{equation}
 f(z_i)=\frac{A_p\sum_{j=1}^{n(z_i)}{I_j}}{\Delta z_i}
\end{equation}
- ``cumulative fraction vertical profile'': derived by summing the pixels values located within $|z|<z_i+\Delta z_i$ and dividing by the total 
intensity on the map $I_{\rm TOT}$, that is:
\begin{equation}
 f(z_i)=\frac{\sum_{j=1}^{N(z_i)}{I_j}}{I_{\rm TOT}}
\end{equation} 
As one can see, the vertical profiles correspond to each one of the radial profiles defined above and are used for the same purposes.

\subsection{Intrinsic galaxy stellar emission, opacity and gas mass distribution}
\label{res_sect_intrinsic}

\begin{figure}
\centering
\includegraphics[scale=0.35]{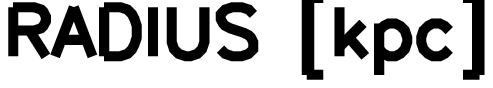}
\includegraphics[scale=0.35]{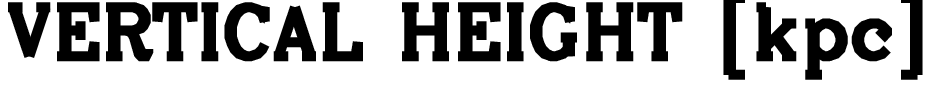}
\caption{Radial (upper panel) and vertical (lower panel) average profiles for the B-band optical depth $\tau_B$ (black continuous line), 
the gas surface density (black dashed line), 
and the intrinsic stellar emission specific intensity I(912 \AA{}) (dark blue), I(2200 \AA{}) (light blue), I(4430 \AA{}) (green) 
and I(2.2$\mu$m) (red). In the upper panel we also show the metallicity average radial profile (black dotted line).
The profiles are normalized by ther maximum value. The normalization factors are shown in 
table \ref{norm_factors_orig}.}
\label{prof_dust_stars_orig}
\end{figure}

\begin{figure*}
\includegraphics[scale=0.5]{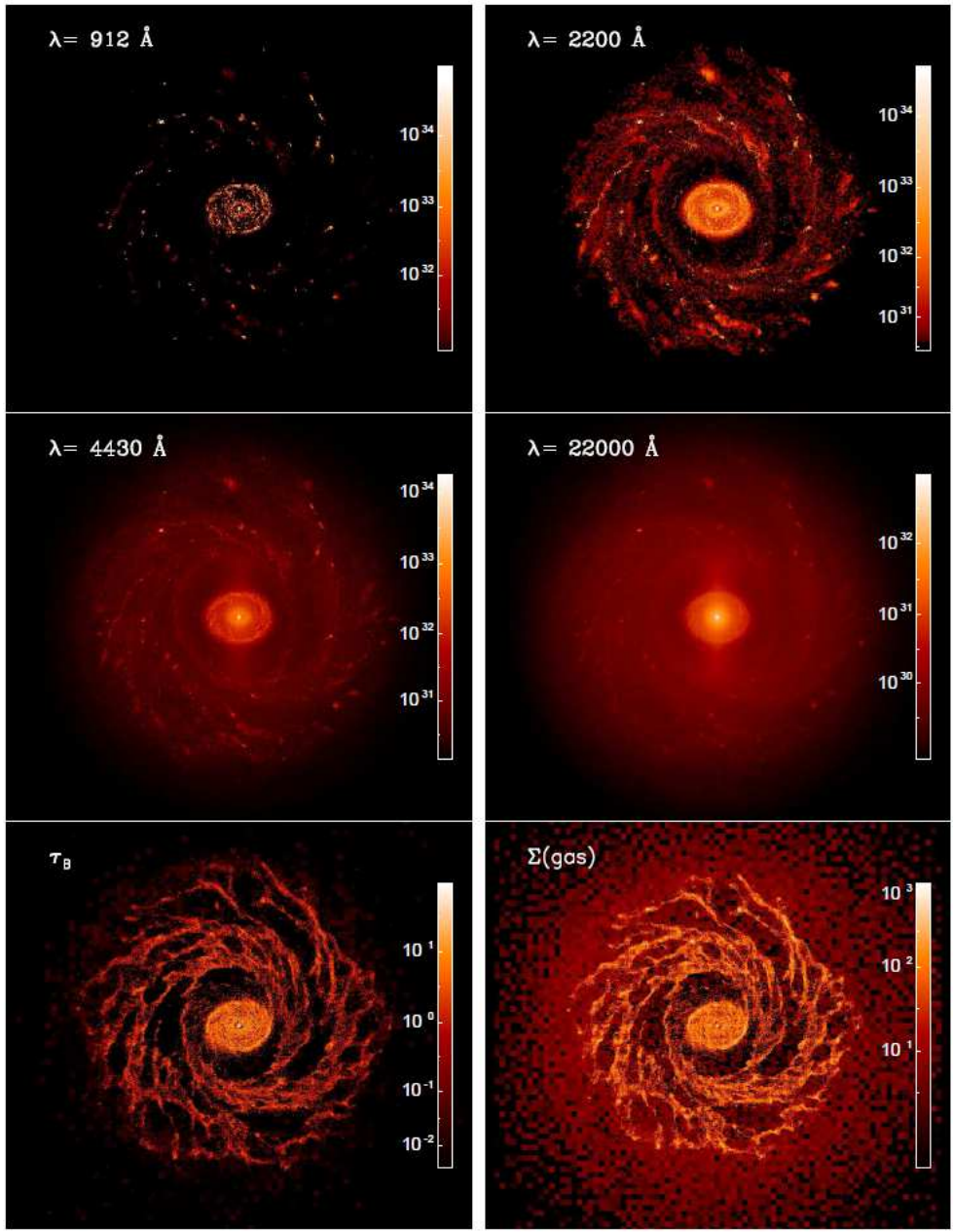}
\caption{Face-on view maps of the galaxy intrinsic stellar light brightness at UV (912 and 2200\,\AA{}), optical (4430\,\AA{}) and 
near-IR (2.2\,$\mu$m) wavelengths, of the optical depth at 4430\,\AA{} (within the B-band) and of the gas mass surface density $\Sigma_{\rm gas}$.   
The units of the values aside of the colour bar are erg/(s\,\AA{}\,sr\,pc$^2$) for the stellar brightness and M$_\odot$/pc$^2$ for the 
gas mass surface density.}
\label{galaxy_face_on_intrinsic}
\end{figure*}

In order to understand the characteristics of the predicted stellar and dust emission maps, it is important to first understand what the 
intrinsic distributions of stellar emission at different wavelengths in the simulation are, as well as the dust opacity and
gas mass distributions. This is also important for comparison with our empirical knowledge of these distributions derived 
from observations of real galaxies. For these reasons we show in Fig. \ref{prof_dust_stars_orig} the 
average radial (upper panel) and vertical (lower panel) profiles of the intrinsic stellar emission specific intensity at 912 \AA{}, 2200 \AA{}, 
4430 \AA{} and 2.2$\mu$m, the dust optical depth at 4430 \AA{} (within the B-band) and the gas mass surface density for the face-on and 
edge-on views. Within the upper panel, we also included the metallicity average radial profile. To compare the profiles in a single 
plot, we normalized each profile to its maximum value. The normalization factors can be found 
in Table \ref{norm_factors_orig}. Furthermore, in Fig. \ref{galaxy_face_on_intrinsic} we show the face-on maps for the surface 
brightness distribution of the intrinsic stellar emission at the same wavelengths as well as the maps of the B-band dust optical depth and 
gas mass surface density distributions. These maps contain essential information to interpret not only the radial profiles 
in Fig.\ref{prof_dust_stars_orig} but also some of the results shown in the next section. 

The face-on radial profiles show the following characteristics. The stellar emission at UV wavelengths (912 and 2200 \AA{}) shows two
distinct regions of emission: a central region (for r $\lesssim$ 2\,kpc), corresponding to the nuclear disc analysed by Cole et al. (2014),  
which is the brightest UV emitting region in the galaxy; a more extended region 
(hereafter also referred to as ``spiral disc''), whose average brightness 
is rather constant in the range 3$<$r$<$5\,kpc but decreases quickly at larger radii. The two regions are separated by a dip in the emission at $\approx$2\,kpc. 
From the maps in Fig.\ref{galaxy_face_on_intrinsic}, 
we see that the extended region outside the nuclear disc appears to be rather clumpy at 912\,\AA{}, with peaks corresponding to young star 
formation regions. Instead, at 2200\,\AA{} a more diffuse flocculent emission component is present as well, apart from the clumpy emission. 
Although the stellar emission at longer wavelengths (4430 \AA{} and 2.2$\mu$m) 
is, like the UV, bright at small radii, the emission profile does not show a clear dip at about 2\,kpc, presenting instead 
 a smooth radial profile which decreases regularly with radial distance. The smooth radial profile at these wavelengths 
is due to the enhanced diffuse stellar emission evident on the optical and near-infrared maps in Fig.\ref{galaxy_face_on_intrinsic}.
The profile of the B-band optical depth, as well as the morphology of the dust optical depth map,
resemble more the UV than the optical/near-infrared stellar emission distributions. The dip at about 2\,kpc is evident as well but it is less 
pronounced compared to the UV one. We also note that a bar, bright at optical/near-infrared wavelengths, is not evident in either the 
UV or the optical depth maps. This is because in this simulation the bar consists of old stellar populations but not of dusty gas and young stars. 
Finally, it is interesting to note that the gas mass surface density profile does not decrease with radius as quickly as the optical depth 
profile. The different behaviour is due to the fact that we assumed a dust-to-gas mass ratio dependent on both the gas metallicity 
and temperature (see section \ref{input3d_sect}). However, as it can be realized by comparing the gas, metallicity and optical depth 
radial profiles in Fig.\ref{prof_dust_stars_orig}, the metallicity scaling is the predominant factor causing the 
difference between the profiles of gas mass surface density and optical depth.
This simulation does not exhibit an extended UV disc, as commonly observed in nearby galaxies (Gil de Paz et al. 2005,  
Thilker et al. 2007). However, galaxies are known to present a wide range of disc radial profile morphologies (Mu\~{n}oz-Mateos et al. 2009). 
In particular, the UV and near-IR profiles of the simulated galaxy resemble qualitatively those found by 
Mu\~{n}oz-Mateos et al. (2009) for M 81 (see their Fig. 6), including the inner and outer ``truncations'' in the UV profile (these features
are present also when the effect of dust is included, see section \ref{res_sect_profiles}).   
As discussed by those authors, within the outer UV truncation radius ($\approx6$\,kpc in our simulation) the galaxy UV to near-IR colour 
become bluer with radius, consistent with the inside-out galaxy growth scenario. Beyond that radial position, the trend in colour is reversed.
This change in colour gradient was predicted as a result of radial migration of old stars to the galaxy outskirts by Ro\v{s}kar et al. (2008) 
and then found by Bakos et al. (2008) for galaxies with truncated discs.

The vertical profile, obtained from the edge-on view maps, shows that the UV emission decreases more quickly with
 distance from the plane than the optical/near-IR emission. This is consistent with the young stellar populations
being more confined within a thin disc, while the old stellar populations smoothly extends to larger heights.
The B-band optical depth profile follows quite closely the 2200 \AA{}  profile. 
The gas mass surface density profile follows the 
optical depth profile until about z=200\,pc but then flattens out instead of decreasing rapidly.  
As for the radial profiles, this difference is due to the varying dust-to-gas mass ratio we assumed when creating the 3D grid. 

From the comparison of both the radial and vertical profiles, we can see that the global distributions of stellar emission, opacity and 
gas distribution in the simulated galaxy contain the main characteristics observed for real galaxies: the gas and the optical depth roughly 
follow the path of star formation (traced by UV emission); the old stellar populations 
(traced by 4430 \AA{} and 2.2$\mu$m) have a smoother radial distribution compared to the young stellar population and a larger vertical height. 

We note that the inferred distribution of gas and metals on sub--kpc scales is not completely realistic, since the galaxy dusty ISM is 
confined only to the flocculent spiral structure. 
To test whether the simulated galaxy contains enough gas, we made a comparison with data points in Fig 2 of Peeples \& Shankar (2011)
showing mainly the HI-to-stellar mass ratio as a function of stellar mass for a sample of local galaxies, and adjusted for the empirical 
calibration of H2/HI from Saintonge et al. (2011). The simulated galaxy has log(M$_{\rm gas}$/M$_{stars}$)=-1.19 which is about 0.4 dex less 
than the median value measured for galaxies with similar stellar mass (that is, log(M$_{\rm stars}$)=10.8). 
This discrepancy is comparable to about 1.5 times the observed rms scatter. From this comparison it seems that the predicted gas 
content is very much on the low side of the observed distribution, though not completely detached from this distribution.

About the simulated galaxy metal content, we note that real galaxies are found to lie on a rather tight stellar mass versus metallicity 
relation (see e.g. Tremonti et al. 2004, Zahid et al. 2014). In order to derive this relation, the metallicity is measured from the intensity 
of strong nebular line emission, while the stellar mass from the total luminosity in a particular band times a suitable mass-to-luminosity
ratio. The metallicity is typically expressed in terms of number of oxygen per hydrogen atoms, that is, in the form 12+log(O/H). 
Instead, in our calculation, we used iron as a tracer of metallicity. Because of this, we obtained a rough estimate of the 
average oxygen abundance by assuming that the iron abundance 
 scales as the oxygen abundance and assuming a solar value of 12+log(O/H)=8.69. In this way, we derived that the average oxygen abundance 
in our galaxy simulation is equal to 8.47 (considering only the volume within 200 pc from the galaxy plane). According to Fig. 6 of 
Tremonti et al. (2004) the median value for the simulated galaxy stellar mass is 9.1, which is about 0.6 dex higher. This indicates 
that the overall metallicity of the simulated galaxy is low compared to the empirical estimate and, thus, highlights a global lack of 
metals in the simulated galaxy disc. However, since the metallicity measurement is derived from nebular emission lines, 
produced mainly in star formation regions, it might be that the comparison with the data is not completely 
fair if we average the metallicity over the entire galaxy disc. If we consider only the galaxy regions within 1 kpc from the 
centre, where most of the galaxy star formation is located, the average metallicity would be
8.85. This value is still 0.2 dex smaller than the empirical median value. If we consider the latter estimate as an upper limit, 
we can conclude that there is an appreciable lack of metals in the galaxy simulation. 

Evidence for a low amount of metals in the simulated galaxy comes also from the comparison of the dust-to-stellar mass ratio 
with the empirical values for galaxies with similar stellar mass. Grootes et al. (2013) considered a sample of local spiral galaxies and 
found that the median value for a stellar mass similar to that of 
the simulated galaxy is about log(M$_{\rm dust}$)=8.2 (see their Fig. 2). This value is about 1.1 dex higher than the simulated galaxy dust 
mass (see Table \ref{table_galaxy param} for the galaxy parameters). 
Similarly, the median value of the dust-to-stellar mass ratio found by Smith et al. (2012) for a spiral galaxy sample in the HRS survey is also 
a factor 10 higher than the simulated galaxy value. The scatter in the empirical values ($\approx$ 0.2 dex) is not large enough to account 
for the discrepancy. Since the dust mass is related to both gas and metal abundance in the ISM, this finding is in principle consistent with a 
non negligible lack of either or both these components in our simulation. In the light of the above discussion, we interpret the low
dust-to-stellar mass ratio of the simulated galaxy as being most likely due to a combination of both a low gas and metal content.

Finally, we note that the global amount of star formation rate ($SFR\sim3~M_\odot/yr$, as derived from the N-body+SPH simulation)
is consistent with empirical values. By comparing the 
intrinsic galaxy luminosity at 2200 \AA{} with the values found by Grootes et al. (2013) for a sample of local spiral galaxies 
(corrected for attenuation), we found that the simulated galaxy UV luminosity is within 0.1-0.15 dex from the empirical median value for 
galaxies of same stellar mass. This is within the scatter in the attenuation corrected L(2200 \AA{}) versus stellar mass relation found 
by Grootes et al. (2013).

\begin{table}
\begin{center}
\caption{Normalization factors for radial and vertical average profiles shown in Fig.\ref{prof_dust_stars_orig}.}
\begin{tabular}{l | c | c | l|}
\hline\hline
Parameter & Radial & Vertical & Units \\
\hline
$\tau_B$ & 10.53  & 81.38 & \\
$\Sigma_{\rm gas}$ & 117 & 1970 & M$_\odot/pc^2$ \\
$Z $ & 0.038 & -- \\
I(912 \AA{}) & 7.57e+33 & 6.44e+33 & erg/(s $pc^2$ sr \AA{}) \\
I(2200 \AA{}) & 4.61e+33 & 3.34e+33 & erg/(s $pc^2$ sr \AA{}) \\
I(4430 \AA{}) & 2.74e+33 & 1.43e+33 & erg/(s $pc^2$ sr \AA{}) \\
I(2.2$\mu$m) & 2.74e+32 & 9.46e+31 & erg/(s $pc^2$ sr \AA{}) \\
\hline
\end{tabular}
\label{norm_factors_orig}
\end{center}
\end{table}

\section{Results}
\label{results_section}
In this section we show how the simulated galaxy is predicted to look at various wavelengths, both in stellar light and dust emission, 
using DART-Ray. In addition, we present an analysis of the attenuation, scattering and heating characteristics of dust in the simulated galaxy
on both global and local scales and for different wavelengths and inclinations.   
Although the simulation is taken to be a realistic representation of a disc galaxy, 
we have to keep in mind that the simulation is still a model
of a galaxy and improvements on the input physics are still underway. However, many of its global characteristics resemble quite well those 
seen in real galaxies (Sect.\ref{res_sect_intrinsic}) and for this reason we perform the analysis mentioned above.  
Note that for many figures in this section we show maps 
and profiles only for a few selected wavelengths. Additional figures for a larger set of wavelengths are shown in the appendix.

\subsection{Global emission SED}
\label{res_sect_glob_sed}

\begin{figure}
\centering
\includegraphics[scale=0.3]{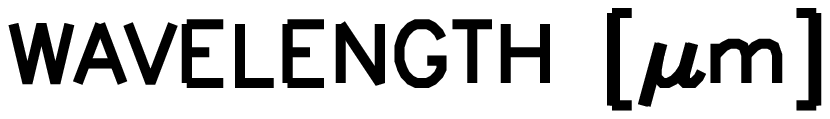}
\caption{Galaxy total stellar and dust emission SEDs. The blue curve is the intrinsic stellar 
luminosity spectra. The red curves are the predicted stellar luminosity spectra at the input inclinations ($\theta$=0, 12.8, 25.7, 38.6, 51.4, 64.3, 77.1, 90\,deg)
from face-on ($\theta$=0\,deg, upper curve) to edge-on ($\theta$=90\,deg,bottom curve). Note that the predicted stellar luminosities have been derived by multiplying by 4$\pi$ the outgoing radiation luminosities per 
unit solid angle predicted at each inclination. The black curve on the right side of the plot represents the total dust/PAH luminosity 
spectra. The wavelength sampling of the stellar SED is as in the table E.2 of P11.}  
\label{tot_sed_vgal}
\end{figure}

In Fig.\ref{tot_sed_vgal} we show the global (spatially integrated) intrinsic SED of stellar emission, the predicted observed
stellar emission SED for the input set of line-of-sight directions (see section \ref{input3d_sect}) and the dust emission SED 
calculated using the adaptive library approach described in section \ref{non_LTE_dust_sect}. Note that the dust emission does not depend on the 
inclination because we have assumed it to be optically thin (this assumption might not be completely correct for light travelling in
the plane of the galaxy at wavelengths in the near/mid-IR part of the dust 
emission SED. However, it is a good approximation for the FIR part, where most of the dust luminosity is emitted). 

By comparing the intrinsic and attenuated stellar emission SEDs in Fig.\ref{tot_sed_vgal}, one can notice that a large fraction of the UV 
radiation is attenuated by dust. Moreover, the attenuation is not negligible in the optical/near-IR. 
The total intrinsic stellar luminosity of the galaxy between 912 \AA{} and 50000 \AA{} is 2.57$\times10^{44}$ erg/s, of which 
33\% is absorbed and then re-emitted by dust in the infrared. This is in the range of observed values for spiral galaxies: e.g.
Popescu \& Tuffs (2002), Driver et al. (2007), Driver et al. (2008). The fact that the dust emission luminosity is within the 
 typically observed values might seem unexpected because of the low dust content of the galaxy (see Section \ref{res_sect_intrinsic}).
However, we note that the galaxy hosts a central nuclear disc particularly luminous in the infrared (approximately 75\% of the 
total dust emission).

\subsection{Predicted surface brightness distributions of stellar light and dust emission}
\label{res_sect_predicted}

\begin{figure*}
\includegraphics[scale=0.5]{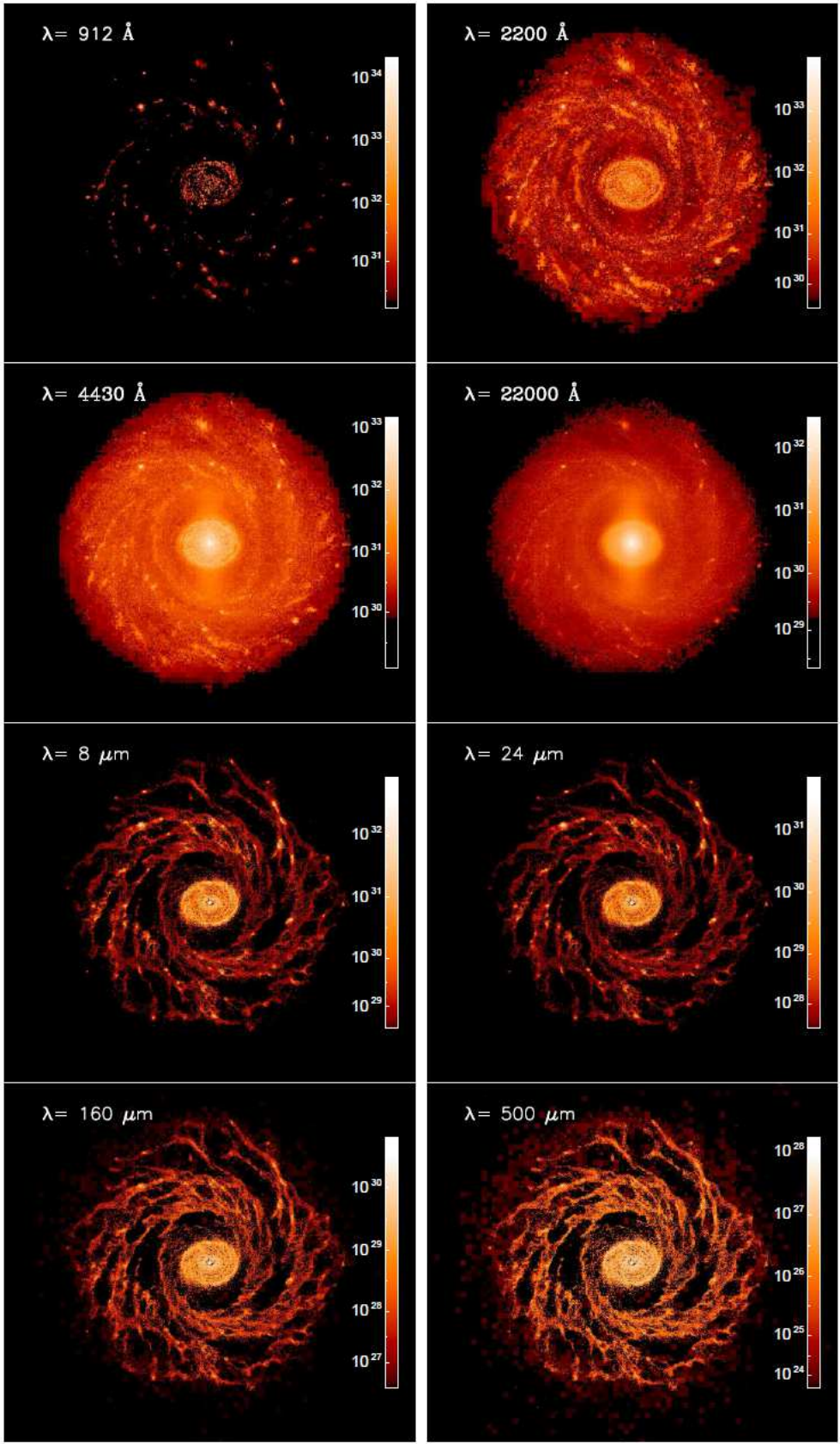}
\caption{Predicted face-on view maps of the galaxy stellar light brightness at UV (912 and 2200\,\AA{}), optical (4430\,\AA{}) and 
near-IR (2.2\,$\mu$m) wavelengths and of the dust emission brightness at a set of infrared wavelengths (8, 24, 160, 500\,$\mu$m). 
The units of the values aside of the colour bar are erg/(s\,\AA{}\,sr\,pc$^2$). Maps for additional wavelengths are shown in the appendix. 
Fig. A1}
\label{galaxy_uv_opt_idir0}
\end{figure*}

Fig.\ref{galaxy_uv_opt_idir0} shows the predicted surface brightness distributions of stellar light and dust-reradiated stellar 
light for the simulated galaxy seen face-on, for a set of UV/optical/near-IR and MIR/FIR wavelengths
The main features of the UV/optical/near-IR maps are as follows:\\
i) the star forming regions and hosting spiral arms are most prominent at
2200\,\AA{}, and less so at very short UV and near-infrared wavelengths. At 912\,\AA{} the
stellar light is observed mainly close to very young star formation regions as in the intrinsic stellar emission map 
(see Fig.\ref{galaxy_face_on_intrinsic});\\
ii) the diffuse light is completely absent at 912\,\AA{}, builds up at 2200\,\AA{}, and
completely dominates in the K band;\\
(iii) the central nuclear disc, hosting substantial star formation, is bright at all wavelengths;\\ 
(iv) as in the intrinsic stellar emission maps, there is a bar which is not visible at short UV wavelengths but it becomes gradually more 
predominant going towards optical/near-IR wavelengths. The increased visibility at longer wavelengths is an intrinsic 
characteristic of the bar, made mainly of old stellar populations, as seen in the intrinsic stellar emission map 
at 2.2\,$\mu$m in Fig.\ref{galaxy_face_on_intrinsic}, and it is not due to dust effects;\\
(v) outside the nuclear disc, the observed emission seems to be dominated by the spiral disc showing a rather flocculent spiral structure.

Dust emission maps, calculated by using the adaptive library approach described in section \ref{non_LTE_dust_sect}, are also shown in 
Fig.\ref{galaxy_uv_opt_idir0} for the face-on view at a set of wavelengths from 8 to 500$\mu$m. 
We chose these infrared wavelengths in order to show PAH emission, emission from  star-forming regions, the peak of the dust emission SED, 
and a wavelength that is sensitive to dust column. The morphology of the dust emission on the maps shows the following characteristics:\\
i) as for the stellar light maps, the most prominent emission comes from the nuclear disc, within which the brightest regions are the galaxy 
centre, and the outer boundary of the disc. In fact, when compared to the spiral disc, the nuclear disc is even more dominant in the dust 
emission than it is in the stellar light. The fundamental reason for this is that the nuclear disk is optically thick (see $\tau_B$ map in 
Fig.\ref{galaxy_face_on_intrinsic}). Furthermore, the nuclear disc is now visible almost like a ring in the dust emission. Again, the
appearance of any ring-like structure containing both dust and stars would indeed be expected to be more pronounced at FIR wavelengths if 
the structure was optically thick, as is the case here;\\
(ii) on the 8 and 24$\mu$m maps, the dust emission in the spiral disc seems to be due to a combination of discrete bright 
compact sources distributed over the flocculent spiral structure. These bright sources correspond to regions of recent star formation  
(as can be seen for example in the distribution of intrinsic stellar emissivity at 2200\,\AA{}, corresponding to star formation over the 
last $\sim$ 100\,Myr, see Fig.\ref{galaxy_face_on_intrinsic}). Some are also visible on the stellar light surface brightness maps, especially 
at UV wavelengths;\\
(iii) the emission is slightly more smoothly distributed on the 160 and 500$\mu$m maps. Overall, the appearance of the galaxy is dominated
by the brightest star formation regions at all IR wavelengths;\\
(iv) the dust emission distribution at 160\,$\mu$m and especially at 500\,$\mu$m resemble quite closely the optical depth distribution shown in 
Fig.\ref{galaxy_face_on_intrinsic}. This shows, as expected, that the submm emission is a tracer of dust column density;\\
(v) there is a general lack of interarm diffuse dust emission. This contrasts with the fact that a smooth interarm emission is commonly 
observed at MIR and FIR wavelengths in nearby galaxies. The reason for this is that the current subgrid metal and mechanical feedback 
implementations in the galaxy simulation generate an interstellar medium without a smooth interarm component which is metal enriched. 
Since we used the gas metallicity to trace dust abundance, in several interarm regions there is simply no dust at those positions; \\
(vi) in contrast to the stellar light maps, there is no evidence of a bar structure on the dust emission maps. This is due to the fact that
the bar predominantly consists of stars and not of dusty gas at time t=10\,Gyr in this simulation.\\

Fig. A3 in the appendix shows the UV and optical/near-IR maps for a line of sight inclined by 51\degree   
to the galaxy rotation axis (hereafter, we will refer to this particular line of sight as the ``inclined view''). 
The dust emission maps for this inclination are also shown in the appendix (Fig. A5).
The morphology of these maps present the same features and trends as seen in the face-on maps. 

\begin{figure*}
\includegraphics[scale=0.5]{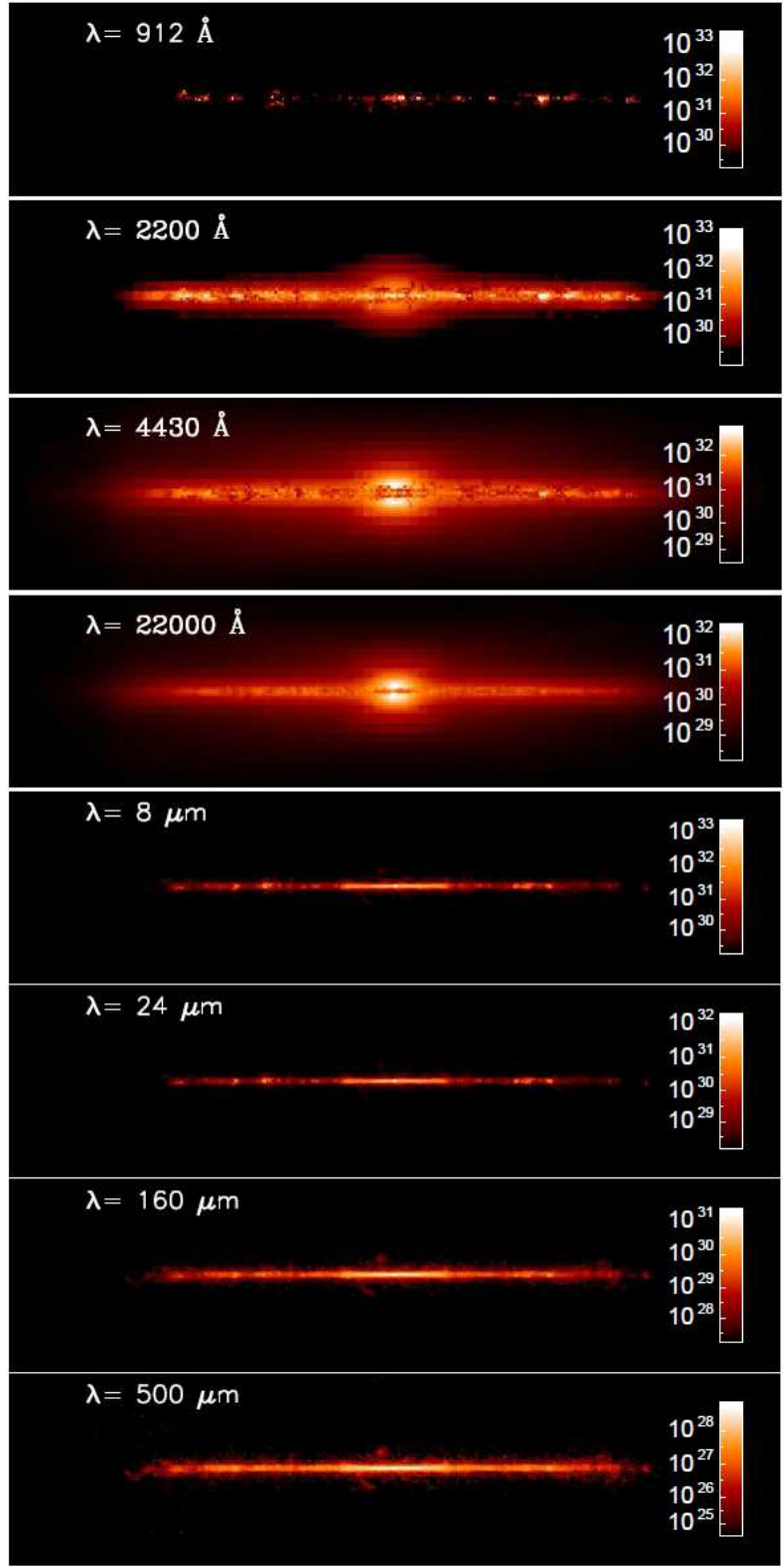}
\caption{Predicted edge-on view maps of the galaxy stellar light brightness at UV (912 and 2200\,\AA{}), optical (4430\,\AA{}) and 
near-IR (2.2\,$\mu$m) wavelengths and of the dust emission brightness at a set of infrared wavelengths (8, 24, 160, 500\,$\mu$m).
The units of the values aside of the colour bar are erg/(s\,\AA{}\,sr\,pc$^2$). Maps for additional wavelengths are shown in the appendix 
(Fig. A6).}
\label{galaxy_uv_opt_idir7}
\end{figure*}

Fig. \ref{galaxy_uv_opt_idir7} shows the predicted edge-on maps at UV, optical and near-IR wavelengths. 
We notice the following features of particular interest:\\  
i) the most obvious property of the maps is the fact that a prominent dust lane is not clearly visible at all wavelengths. The only exception 
is a dust lane appearing between 4430 and 22000\,\AA{} (see also Fig. A6 in the Appendix). This is 
confined to the central regions seen against the light of the bulge, in contrast to the 
appearance of typical spiral galaxies seen edge-on, which show a prominent dust lane in the optical/NIR bands through the whole disc 
(e.g. Xilouris et al. 1999, De Geyter et al. 2014). The absence of a prominent dust lane in the optical/NIR cannot be ascribed to resolution effects, and it is likely 
due to the absence of a diffuse dust component, as already noted in connection with the absence of a diffuse infrared emission in the spiral 
disc;\\
(ii) in contrast to the optical/near-IR, at UV wavelengths the dust lane is completely absent. This is due to the fact that in the edge-on 
view the UV emission is optically thick within the disc, except for its external part facing the ``observer''. Therefore, in the UV we are 
seeing mainly the emission from stars located at large radii on the near side of the galaxy;  \\
(iii) the central dust lane, visible in the optical and near-IR, becomes thinner going to longer wavelengths. 
This effect is probably due to the different edge-on optical depths observed at the different wavelengths (the optical depth is smaller 
at longer wavelengths) as well as differences in the intrinsic vertical distribution of stellar emission;\\
(iv) the stellar light observed at short UV wavelengths (912-1650 \AA{}) originates predominantly from the regions close to the galaxy plane; \\
(v) from 2000\,\AA{} to longer wavelengths in the UV, also the bulge become visible. From an inspection of the intrinsic stellar 
emission edge-on map, we notice that the increasing visibility of the bulge is not due to the presence of dust but it is an intrinsic feature 
of the galaxy stellar emission;\\
(vi) although one sees light in the UV coming from the galaxy halo, this is not as prominent as seen in nearby edge-on 
galaxies in the local universe where the UV halo emission is comparable to or even exceeds the direct light seen from the disc 
(Seon et al. 2014, Hodges-Kluck \& Bregman 2014). 

From the edge on view maps of the dust emission, also shown in Fig.\ref{galaxy_uv_opt_idir7}, one can notice that:\\
(i) at all infrared wavelengths the disk is thinner than in the UV/optical. This is to be expected as the ISM in the disk cools radiatively, 
 so will settle into a configuration with smaller scale height than the stars;\\
(ii) no bulge is seen in dust emission, as expected since the bulge does not host a cold ISM; \\
(iii) some extraplanar dust emission is evident from the map as elongated emission extending from the galaxy plane, especially at the longest
wavelengths. However, no clear extraplanar diffuse dust emission is evident from the maps. In contrast to the stellar light 
optical/near-infrared edge-on maps, the dust emission morphology does not present a clear emission from the galaxy halo.

\subsection{The effect of dust on the average radial profiles of the stellar light surface brightness}
\label{res_sect_profiles}
One of the effects of the presence of dust on the galaxy appearance is the modification of the observed profiles of stellar emission
compared to the intrinsic ones. This effect has been extensively studied, both observationally (e.g. Boissier et al. 2004, 
Popescu et al. 2005) and theoretically. However, the theoretical studies have been exclusively based on
axi-symmetric models  (Cunow 1998, Cunow 2001,  Byun et al. 1994, Kuchinski et al. 1998,  Pierini et al. 2004, Tuffs et al. 2004, 
M\"{o}llenhoff et al. 2006, Gadotti et al. 2010, Pastrav et al. 2013a,b).
It is therefore interesting to see how the predictions change when more complex structure is present. 

\begin{figure}
\centering
\includegraphics[scale=0.38]{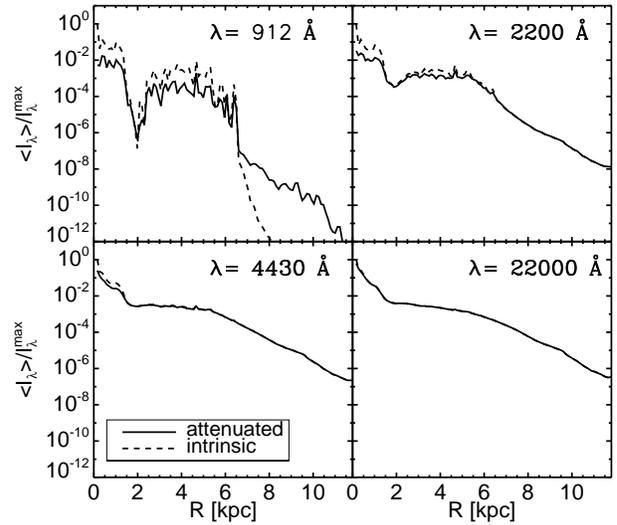}
\caption{Comparison of the average radial profiles for the intrinsic (dashed line) and attenuated (continuous line) galaxy 
emission brightness for the face-on view. The profiles have been normalized to the maximum value of the intrinsic profile. 
Profiles for additional wavelengths are shown in the appendix (Fig. A2).} 
\label{galaxy_comp_stars_norm_prof_idir0}
\end{figure}

In Fig.\ref{galaxy_comp_stars_norm_prof_idir0} we show a comparison between the average radial profile
of the ``attenuated'' (that is including the effect of dust) and intrinsic stellar light surface brightness at different wavelengths. 
The profiles have been normalized by the maximum value of the intrinsic profile. We notice the following effects:\\
(i) at UV wavelengths the intrinsic profiles present a steep gradient of emission very close to the galaxy centre ($R<$100-200\,pc). 
This feature disappears in the attenuated profiles; \\
(ii) the dip between the nuclear disc and the extended spiral disc at $\approx$2\,kpc, more evident at UV wavelengths, is less pronounced in 
the attenuated profiles; \\
(iii) at all wavelengths the attenuated profiles decrease less steeply with radius within the nuclear disc, while the steepness
is similar for the spiral disc (except for the 912 \AA{} map where the intrinsic profile decreases much more rapidly at large radii.
The smoother decline of the attenuated profile at this wavelength is due to the presence of scattered light in the outer disk, see section
\ref{res_sect_glob_sca}).

We also compared the attenuated and intrinsic radial profiles derived for the inclined view. We did not find particular 
features which have not been seen already for the face-on maps, except the fact that the 
inner 100-200\,pc do not show the steep gradient for the intrinsic stellar emission at UV wavelengths (see Fig. A4 in the appendix).  

In Fig.\ref{galaxy_comp_stars_norm_prof_idir7} we show the normalized average vertical profiles for the attenuated and intrinsic stellar 
emission. The following particular features are evident from these plots:\\
(i) the primary effect of dust is to increase the relative brightness of stellar emission at higher distance above the plane with respect to 
positions closer to the plane; \\
(ii) for the optical/near-IR bands the maximum of the attenuated profiles is not located at z=0\,pc but at z$\approx$100-150\,pc 
(see also plots in (Fig. A7) of the appendix for additional wavelengths). The smaller values of the attenuated profiles close to the galaxy plane is 
consistent with the central dust lane seen at these wavelengths (see Fig.\ref{galaxy_uv_opt_idir7}).

\begin{figure}
\centering
\includegraphics[scale=0.38]{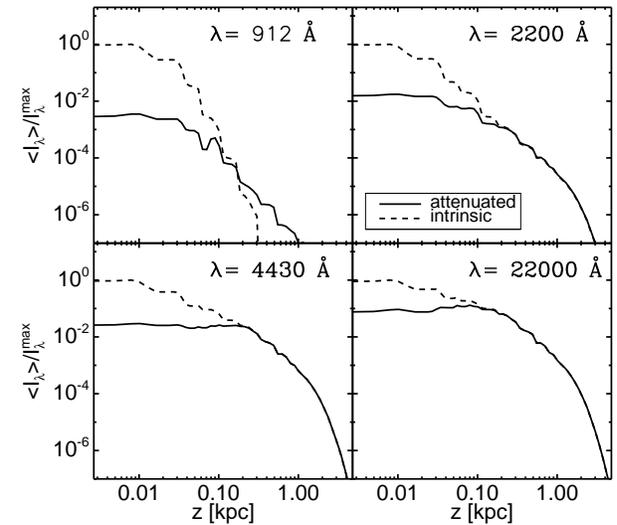}
\caption{Comparison of the average vertical profiles for the intrinsic (dashed line) and attenuated (continuous line) galaxy 
emission brightness for the edge-on view. The profiles have been normalized to the maximum value of the intrinsic profile.  
Profiles for additional wavelengths are shown in the appendix (Fig. A7).} 
\label{galaxy_comp_stars_norm_prof_idir7}
\end{figure}

\subsection{Global stellar light attenuation}
\label{res_sect_glob_att}

\begin{figure*}
\includegraphics[scale=0.32]{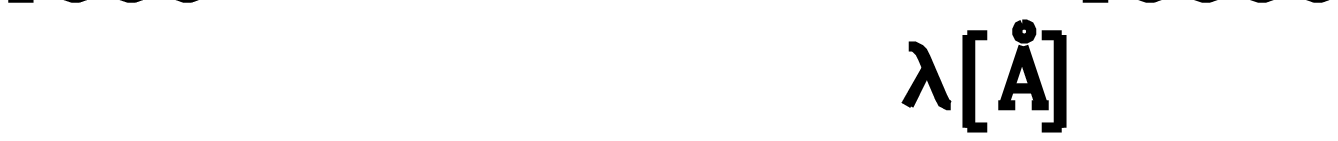}
\includegraphics[scale=0.32]{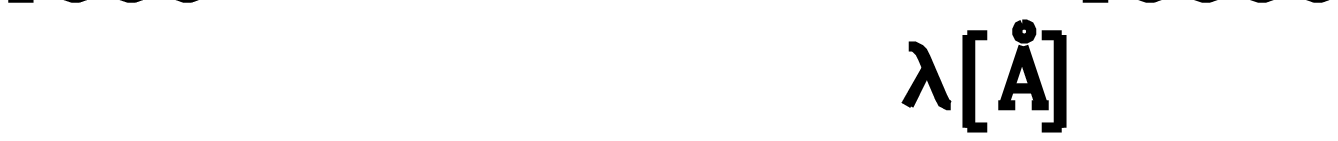}
\caption{Attenuation of the total stellar emission as a function of wavelength. Left panel: the different black curves represent the 
total attenuation for different inclinations 
(($\theta$=0, 12.8, 25.7, 38.6, 51.4, 64.3, 77.1, 90\,deg) from face-on ($\theta$=0\,deg, bottom curve) to edge-on ($\theta$=90\,deg, 
top curve). Right panel: the black curves represent the attenuation normalized to its value at 2.2\,$\mu$m for the same set of inclinations. 
Note that the order of the curves is reversed compared to the left panel (bottom curve=edge-on and top curve=face-on). The red curve is the 
normalised extinction curve from the dust model of Draine \& Li (2007). The green curve represents the normalized attenuation curve of the 
axi-symmetric galaxy model of P11, for parameters of the simulated galaxy 
and for inclination $\theta=60$\degree (see text for more details). The wavelength sampling of the attenuation curve is as in table E.2
of P11.}  
\label{att_vs_lambda_vgal}
\end{figure*}

The plot in the left panel of Fig.\ref{att_vs_lambda_vgal} shows the attenuation curves for the integrated stellar emission at all the 
calculated inclinations. As one can see, the edge-on model is 
the one that presents the highest attenuation (top curve) and the face-on model the lowest (bottom curve), as expected. 
In the right panel of Fig.\ref{att_vs_lambda_vgal} we also show the normalised attenuation curves,
where the normalisation is with respect to the value of attenuation at
2.2\,${\mu}$m (within the K-band range). 

The first thing to observe is that
the normalised attenuation curves appear in reverse order as compared to the
non-normalised ones from the left panel, with the edge-on inclination being the
bottom curve and the face-on inclination being the top curve. For comparison
purposes, we overplotted the dust model normalised extinction curve
(red curve). One can see that the attenuation curves are overall flatter than the extinction curve, with the flatness increasing towards 
higher inclination. This is due to there being a distribution of optically thin and thick lines of sight through the galaxy. 
Whereas the optically 
thin structures might be expected to get dimmer approximately following the extinction law, the attenuation of the optically
thick components will be relatively constant with wavelength.  
Since at high inclination there is a large proportion of optically thick lines 
of sight, the attenuation curves are flatter at the higher inclination. This is a nice illustration of the fact that it is the relative 
distribution of stars and dust rather than the optical properties of the grains that are primarily responsible
for the shape and slope of the attenuation law. 

This conclusion is also in general agreement with the predictions of the axi-symmetric models. To illustrate this we overplotted with the 
green line the normalised attenuation curve predicted by the model of P11, for which the stars and dust in the galaxy are 
distributed in exponential discs, with scale lengths and heights empirically constrained through model fits to images of highly resolved 
edge-on galaxies. We only consider the attenuation from the diffuse component from P11, since in the simulated galaxy the star-forming 
regions are not resolved, and therefore not 
explicitly treated in the transfer of radiation. The plotted curve corresponds to an inclination of 60\degree, a face-on 
central optical depth $\tau^f_B$=4 (approximately the average face-on optical depth of the galaxy within the inner 2\,kpc), 
and a bulge-to-disc ratio B/D=0.1. One can clearly see that, as in the case of the galaxy simulated in 3D, 
the attenuation curve for the 2D model is flatter than the extinction law.
Also, the bump at $\approx$2200\,\AA{}, a particular feature of
the dust model present in the extinction curve, is not as prominent in the
attenuation curves, which show a much smoother enhancement in the same
wavelength region\footnote{As mentioned in Sect.\ref{input3d_sect}, the wavelength grid is taken from table E.2. of
P11. This wavelength scheme was chosen to coursely sample the dust extinction curve for the wavelength range where the curve is smooth, 
and to finely sample the curve in the vicinity of features. Thus, to reproduce the 2175 bump, our
wavelength sampling includes the 2000, 2200 and 2500 A wavelengths.}. This is a well known effect, first shown and discussed in Cimmati et al. (1997) (see also 
Granato et al. 2000, Panuzzo et al. 2007, Pierini et al. 2004, Tuffs et al. 2004, Fischera \& Dopita 2011), and is also due to the 
combined effect of the dust-stellar emission geometry, which make the attenuation curve for the total emission not 
reproducible in terms of a simple foreground screen dust distribution. 

We note that although the stellar light is globally attenuated, this does not exclude that in some localized regions
it can actually be enhanced because of the effect of scattering. We will examine local effects in the following subsection.  

\subsection{Stellar light attenuation on local scales}
\label{res_sect_loc_att}
 
\begin{figure*}
\includegraphics[scale=0.5]{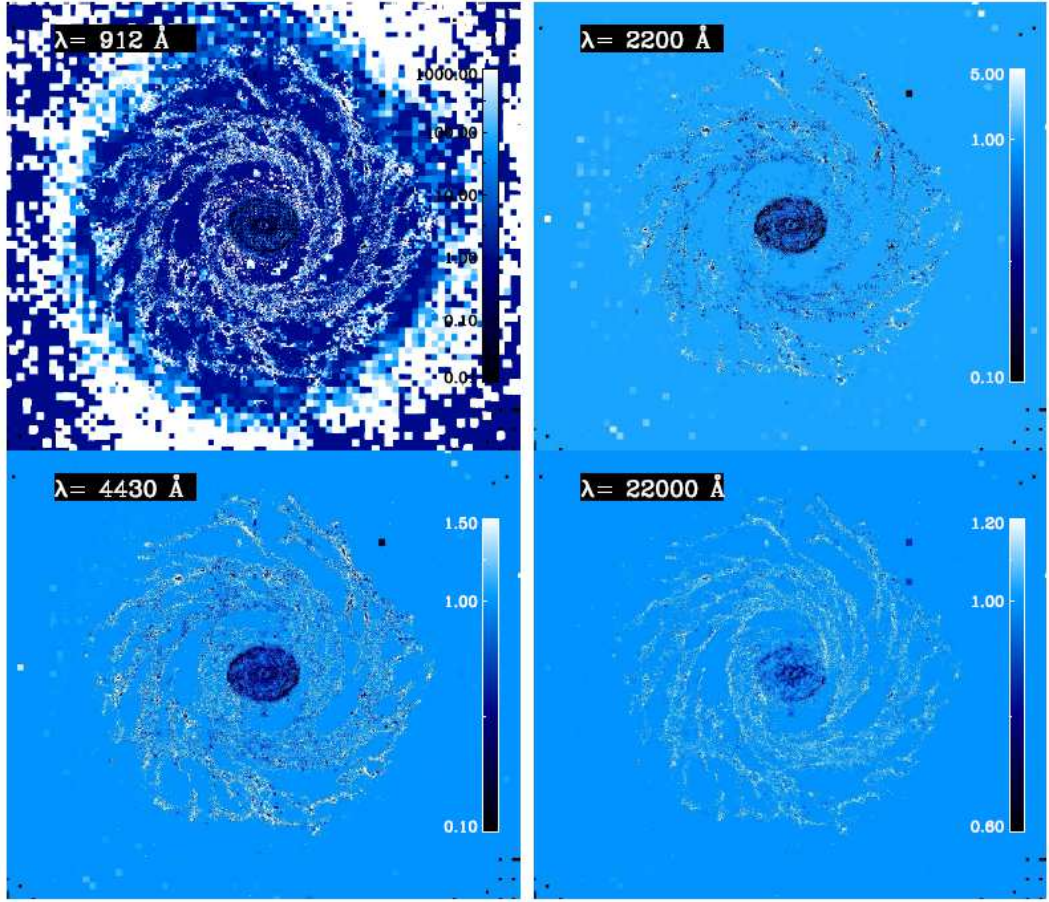}
\caption{Face-on attenuation maps. The pixel values are equal to $I_\lambda/I_{o,\lambda}$, where $I_{o,\lambda}$ and 
$I_\lambda$ are the specific intensities of the intrinsic and predicted emission respectively. Maps for additional wavelengths are 
shown in the appendix (Fig. B1).}  
\label{fig_att_idir0}
\end{figure*}

The panels in Fig.\ref{fig_att_idir0} show the attenuation maps for each wavelength which have been obtained by calculating for each pixel the 
intensity ratio $I_\lambda/I_{o,\lambda}$, where $I_{o,\lambda}$ and 
$I_\lambda$ are the specific intensities of the pixels in the intrinsic and predicted maps respectively. 
In each panel a colour bar shows the colour scales used to define the corresponding range of attenuations (note that
the value ranges and colour scales are generally different for each panel). 
For the face-on attenuation maps we notice the following features: \\
(i) the nuclear disc is highly attenuated at all wavelengths; \\
(ii) there are some regions with $I_\lambda/I_{o,\lambda}>1$, that is, the observed emission is higher than the intrinsic one.
In particular in the FUV some regions far away from the galaxy do not contain many young stars but ``emit'' UV mainly because of scattered 
light;\\ 
(iii) the large scale spiral structure shows a mixed behaviour with some regions which are attenuated and others showing more light than 
the intrinsic local value. 

\begin{figure}
\centering
\includegraphics[scale=0.35]{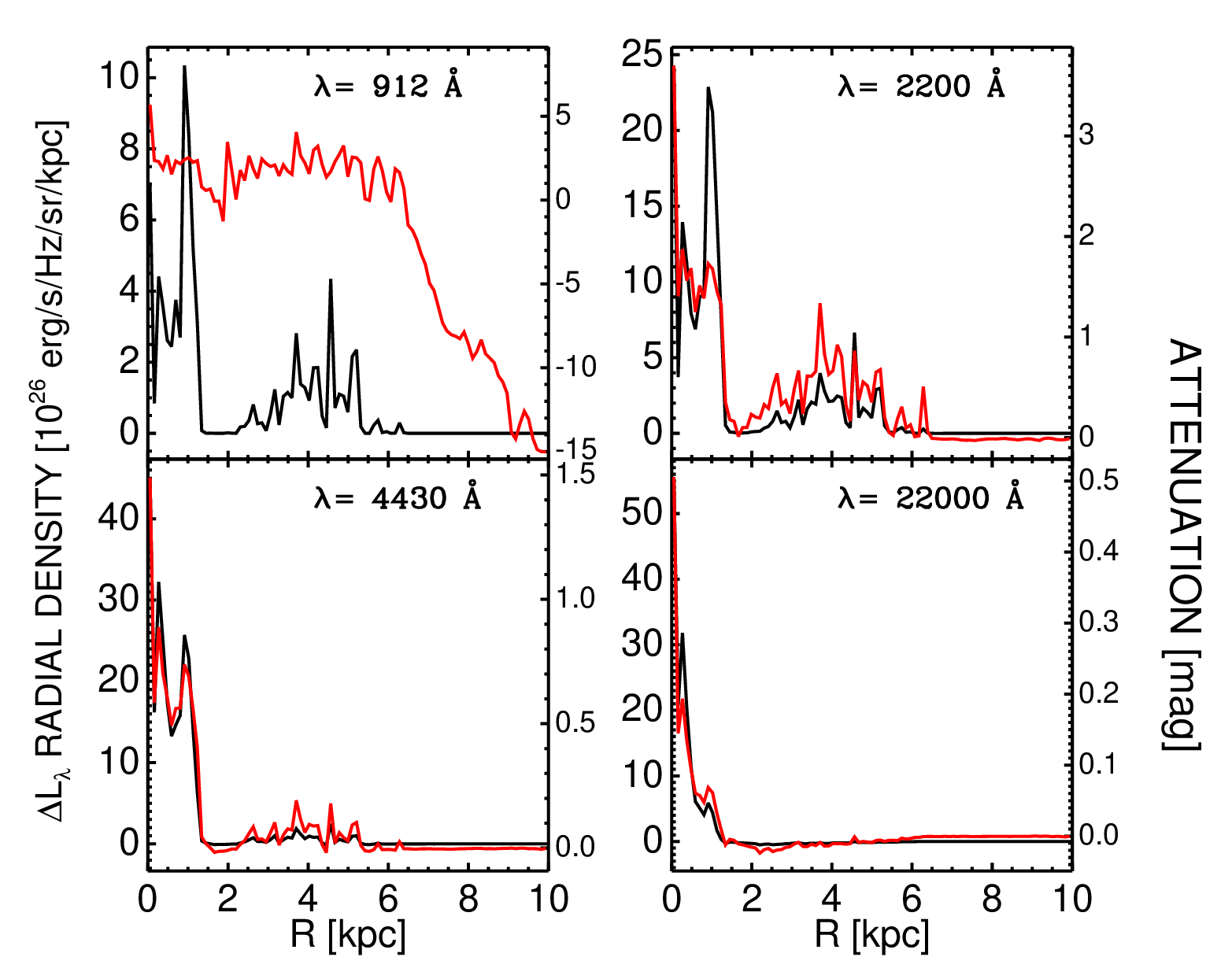}

\caption{Attenuation radial profile (red curve) and $\Delta L_\lambda$ radial density profile (black curve) for the face-on view. 
Profiles for additional wavelengths are shown in the appendix (Fig. B2).}  
\label{prof_att_idir0}
\end{figure}

\begin{figure}
\centering
\includegraphics[scale=0.35]{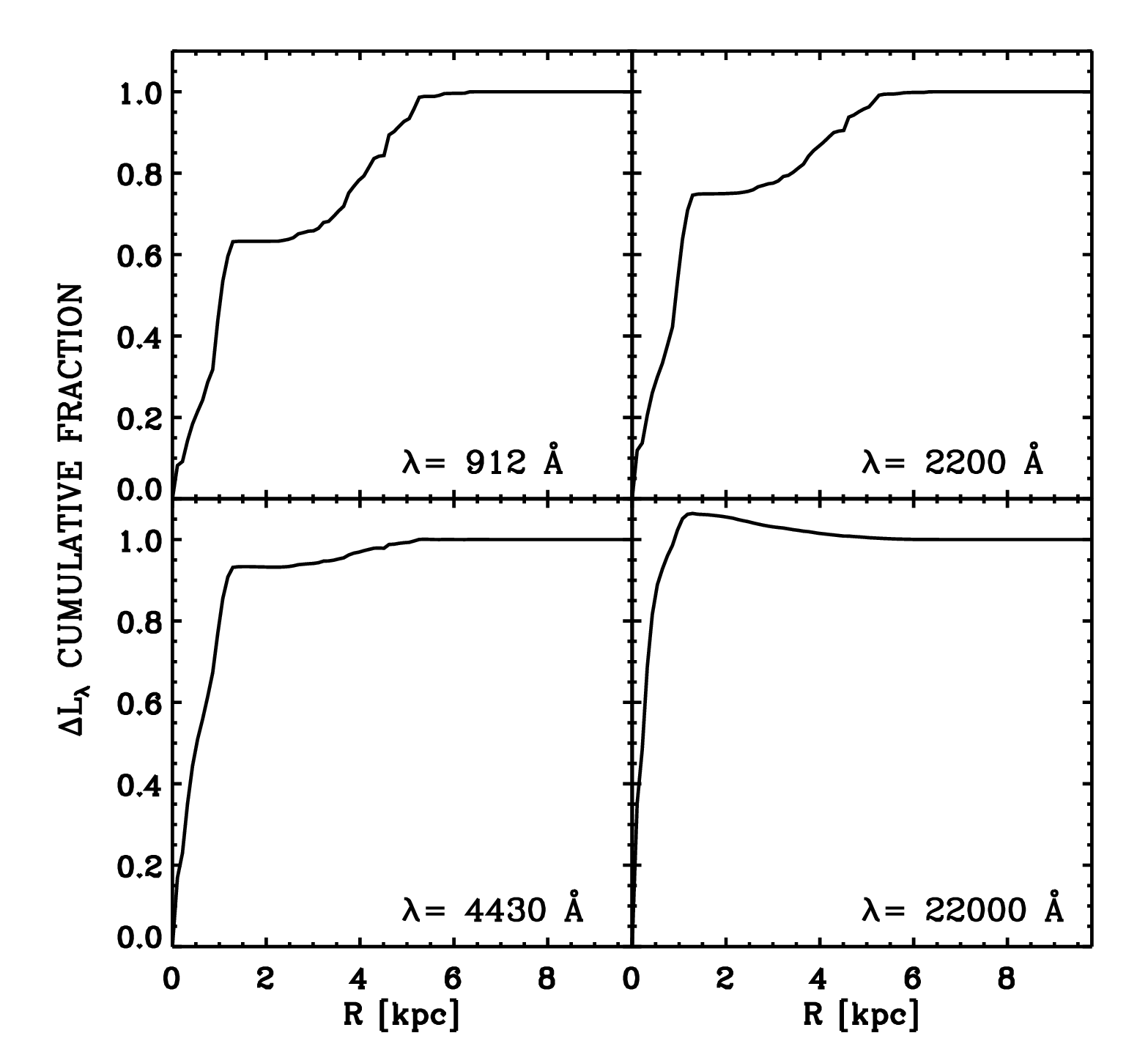}

\caption{Cumulative fraction radial profile of $\Delta L_\lambda$ for the face-on view. Profiles for additional wavelengths are shown in 
the appendix (Fig. B3).}  
\label{prof_att_frac_idir0}
\end{figure}

In order to understand how the attenuation varies radially within the galaxy, we show in 
Fig.\ref{prof_att_idir0} two curves: 1) the radial density profile (see section \ref{sect_def_prof}) of the 
stellar luminosity difference $\Delta L_{\lambda}(R)=L_{\lambda,o}(R)-L_{\lambda}(R)$, where $L_{\lambda,o}(R)$ and $L_{\lambda}(R)$ are 
respectively the intrinsic and attenuated stellar luminosities within each radial bin; 
2) the radial profile of the attenuation $A_\lambda(R)$, derived by averaging radially $I_\lambda$ and $I_{o,\lambda}$ and then calculating 
$A_\lambda(R)=-2.5\log\left(<I_\lambda(R)>/<I_{o,\lambda}(R)>\right)$. 
Through the former curve we can see how much the galaxy is attenuating or enhancing radiation 
at each radii in absolute terms and through the latter we can see how much 
attenuation is important in relative terms in modifying locally the predicted emission at the same positions.  
In addition, in order to quantify the importance of the different radial sectors to the total attenuation,
we also plotted in Fig.\ref{prof_att_frac_idir0} the cumulative fraction of $\Delta L_\lambda$ as a function of radius, calculated as  
 $\int_0^R \Delta L_\lambda(R')dR' / \int_0^{15 kpc}\Delta L_\lambda(R')dR'$.
From these plots we notice the following features: \\
(i) at UV wavelengths the attenuation $A_\lambda(R)$ is higher in the nuclear disc and lower in the spiral disc. An exception 
is the map at 912 \AA{}, where it is almost constant throughout the galaxy until 6\,kpc. At larger radii $A_\lambda(R)$ 
becomes negative in the UV, consistent with the enhanced emission due to scattering at those positions. However,    
 this does not affect much the total attenuation since $\Delta L_\lambda(R)$ is very small at those positions;\\
(ii) the spiral disc is less and less attenuated going towards longer wavelengths. For wavelengths $\geq$8090 \AA{} the 
spiral disc shows negative attenuation\footnote{When using the expression ``negative attenuation'', we will always refer to the attenuation 
expressed in magnitudes.} in the range 2-6\,kpc where the module of $\Delta L_\lambda(R)$, although relatively small, is not negligible. This 
results in a extra contribution to the outgoing radiation which decreases substantially the total amount of attenuation at those wavelengths 
(see below). Note that, although the average attenuation becomes negative at radii R$>$5.5\,kpc for wavelengths between 2200 \AA{} and 4430 \AA{},
the absolute amount of extra luminosity going towards the observer is negligible in those cases and thus this effect does not reduce much the 
total attenuation;\\
(iii) as shown in Fig.\ref{prof_att_frac_idir0}, the cumulative fraction of $\Delta L_\lambda$ due to the nuclear disc goes 
from 60\% at 912 \AA{} up to 100\% at 5650 \AA{} (V band). At longer wavelengths the fraction due to the 
nuclear disc is actually even slightly higher than 100\% but the excess is compensated by the negative attenuation of the larger spiral 
disc. The stellar emission in the near-IR bands is up to 20\% less attenuated thanks to negative attenuation in the spiral disc. 
The importance of scattering for the enhancement of the predicted near-infrared radiation in particular for the face-on view has been already 
pointed out by several authors (Pierini et al. 2004, Baes \& Dejonghe 2001, Tuffs et al. 2004), although not for geometries including a 
complex clumpy structure, as we present here. Here we show that this effect can be important over a significant area of the disk.

We repeated the same analysis for the inclined view (see Fig. B4-B6 in the appendix). In this case, the attenuation maps, the attenuation radial 
profiles as well as the density and cumulative profiles of $\Delta L_\lambda$ show approximately the same features as for the face-on case. 
The only noticeable difference is that there is no substantial negative 
attenuation evident in the spiral disc on the near-IR wavelength maps. In this case, this effect seems to be negligible. 

\begin{figure*}
\includegraphics[scale=0.5]{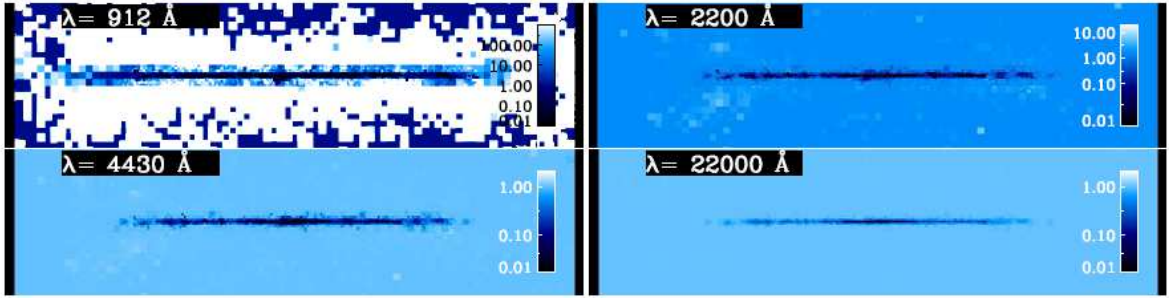}
\caption{Edge-on attenuation maps. The pixel values are equal to $I_\lambda/I_{o,\lambda}$, where $I_{o,\lambda}$ and 
$I_\lambda$ are the specific intensities of the intrinsic and predicted emission respectively. Maps for additional wavelengths are 
shown in the appendix (Fig. B7).}  
\label{fig_att_idir7}
\end{figure*}

\begin{figure}
\includegraphics[scale=0.35]{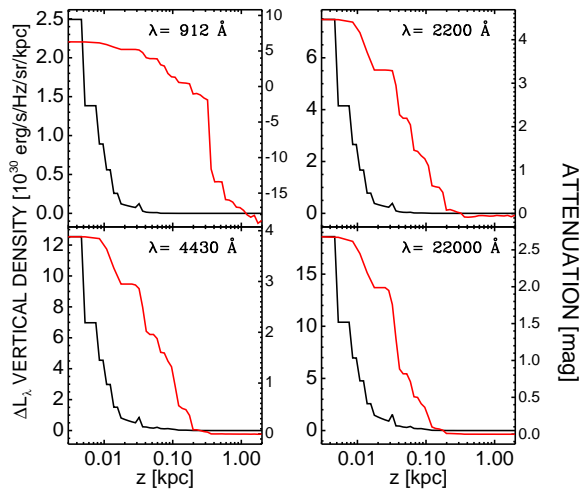}
\caption{Attenuation vertical profile (red curve) and $\Delta L_\lambda$ vertical density profile (black curve) for the edge-on view. 
Profiles for additional wavelengths are shown in the appendix (Fig. B8).}  
\label{prof_att_idir7}
\end{figure}

The panels in Fig.\ref{fig_att_idir7} and \ref{prof_att_idir7} show the attenuation 
maps for the edge-on view, the vertical profiles for the attenuation $A_\lambda(z)=-2.5\log\left(<I_\lambda(z)>/<I_{o,\lambda}(z)>\right)$
 and the vertical density profiles of $\Delta L_\lambda(z) = L_{\lambda,o}(z)-L_\lambda(z)$ (defined in an analogous way as for the 
 radial profiles and derived as described in Section \ref{sect_def_prof}). The most important features that can be noted are the following:\\
(i) the average attenuation is highest at the position of the galaxy plane and decreases quickly going towards 
larger distances; \\
(ii) the attenuation turns significantly negative at around $z=$300\,pc for the UV bands from 912 \AA{} until 2200 \AA{}; \\
(iii) most of the luminosity is attenuated within $z=$100\,pc for the UV bands until 3650 \AA{}. For longer wavelengths the 
attenuation is completed at about $z=$200\,pc.

\subsection{Scattering fraction}
\label{res_sect_glob_sca}

\begin{figure}
\centering
\includegraphics[scale=0.32]{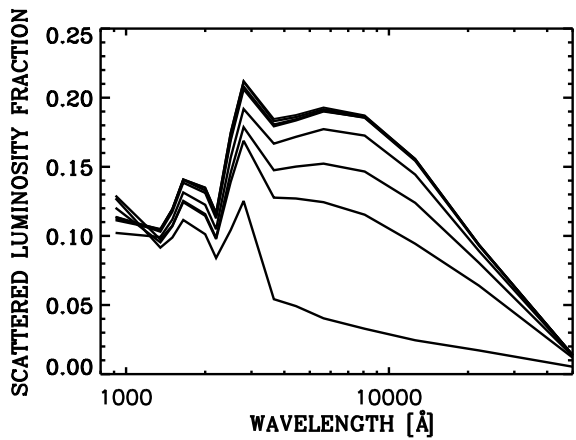}
\caption{Fraction of scattered to total predicted stellar emission as a function of wavelength.
The different curves represent the scattering fraction for different galaxy inclinations ($\theta$=0, 12.8, 25.7, 38.6, 51.4, 64.3, 77.1, 90\,deg)
from face-on ($\theta$=0\,deg, top curve) to edge-on ($\theta$=90\,deg, bottom curve).}   
\label{scatt_frac_vs_lambda_vgal}
\end{figure}

In Fig. \ref{scatt_frac_vs_lambda_vgal} we show the fraction of total observed radiation luminosity produced by stellar light 
which has been scattered at least once before escaping the galaxy towards the observer line of sight. This fraction is plotted as a 
function of wavelength and 
for different inclinations. Apart from the very short UV wavelengths (912 \AA{} to 1500 \AA{}), 
the fraction of scattered luminosity is generally decreasing while 
going to higher inclinations. For a given inclination, the fraction is higher in the NUV/optical regime (that is, between 2200 \AA{} 
and 1.2$\mu$m) and lower in the FUV and NIR (that is, for $\lambda<$2200 \AA{} and $\lambda>$1.2$\mu$m).
An exception is the edge-on case where the UV regime shows the highest fraction of scattered light.   
In any case, it seems that the fraction of scattered luminosity is at most of order of 20\% of the total observed luminosity. 
This means that most of the stellar radiation in the 
predicted surface brightness maps for all the inclinations is due to direct light. We note that our results are in qualitative agreement with those found 
by Pierini et al. (2004) for their set of galaxy models. 

\begin{figure*}
\includegraphics[scale=0.5]{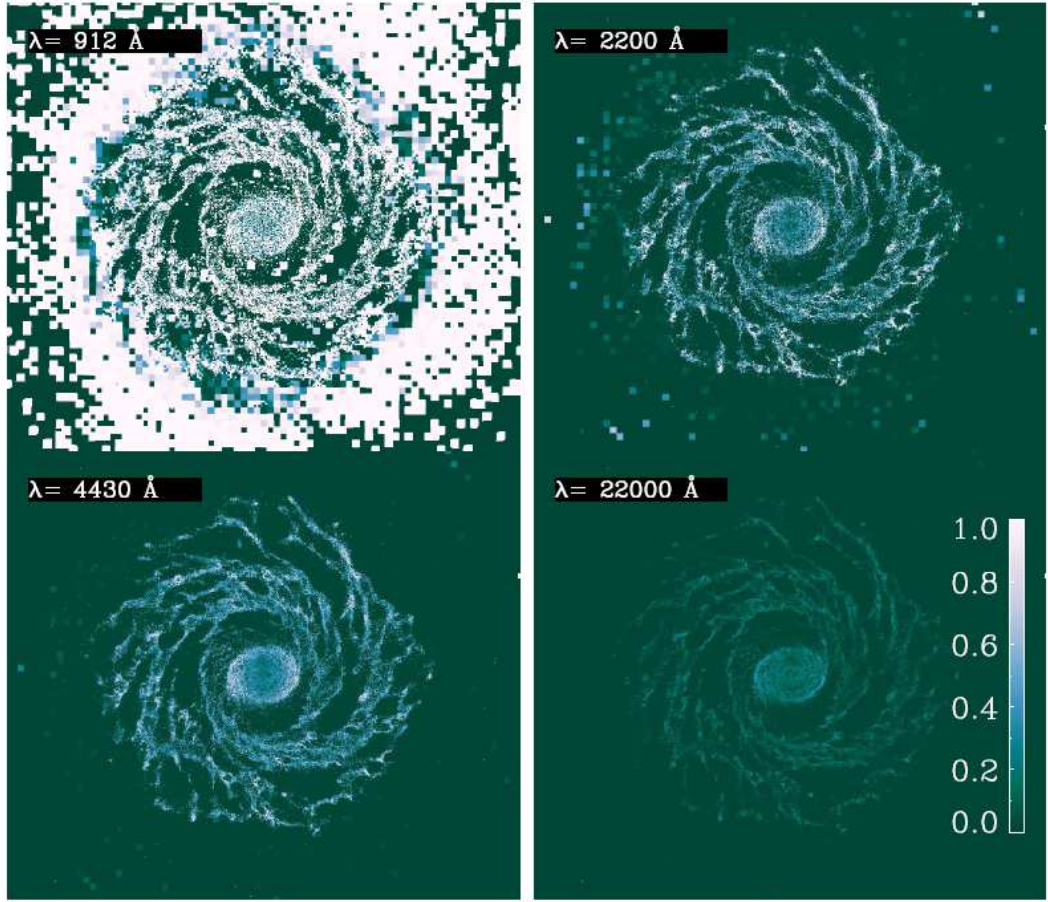}
\caption{Face-on scattering fraction maps. Each pixel value corresponds to the local fraction of observed emission due to scattered radiation.
Maps for additional wavelengths are shown in the appendix (Fig. C1).}  
\label{fig_sca_idir0}
\end{figure*}

Similarly as in the local attenuation section, we present here and in the appendix maps and plots showing the variation of the 
scattered light fraction at different inclinations. Fig.\ref{fig_sca_idir0} shows face-on galaxy maps whose pixel values 
are equal to the fraction of predicted emission due to scattered light at each position for different wavelengths.  
Some features one can notice from these maps:\\
(i) for FUV wavelengths, there are regions at large radii with a high fraction of scattered light. This is consistent with 
the negative attenuation observed at large radii for those maps (see section \ref{res_sect_loc_att});\\
(ii) both the nuclear disc and the spiral disc present a combination of regions where the light is predominantly
scattered or direct; \\
(iii) in some of the interarm regions there is no scattered light. This is because there is essentially no 
dust in those regions in the simulation as remarked in section \ref{res_sect_predicted};\\    
(iv) the external part of the nuclear disc shows higher scattering fractions than the inner part. This coincides with 
the predominance of young stellar populations in heating the dust at the same positions (see section \ref{res_sect_dust_heat}); \\
(v) there is a general decrease of regions with high scattering fraction going towards longer wavelengths. 

We also examined the scattering fraction maps for the inclined view but we did not find major differences compared to 
the face-on view (see Fig. C4 in the appendix).

\begin{figure*}
\includegraphics[scale=0.5]{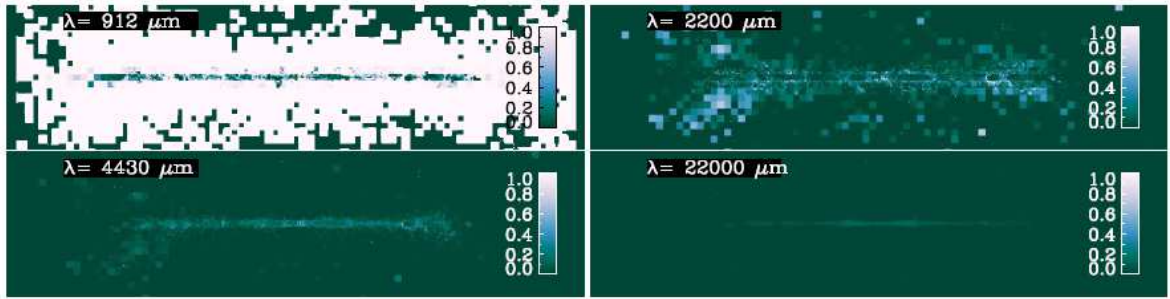}
\caption{Edge-on scattering fraction maps. Same description as in Fig.\ref{fig_sca_idir0}. Maps for additional wavelengths are 
shown in the appendix (Fig. C7)}  
\label{fig_sca_idir7}
\end{figure*}

Finally, we show the scattering fraction maps for the edge-on view in Fig.\ref{fig_sca_idir7}.
From these maps we notice in particular that the scattering fraction becomes very high in the UV at large distances from the galaxy plane. 
This is consistent with the negative attenuation we saw before in section \ref{res_sect_loc_att}. 
It is apparent that the appearance of egde-on galaxies in the UV will be very sensitive to the amount of grains present in the halo 
above the plane of the galaxy. Scattered light radial and vertical profiles are shown and described in the appendix (Fig. C2-C9).  

\clearpage
\clearpage

\subsection{Dust heating}
\label{res_sect_dust_heat}

In this section we present the results for the fraction of dust heating due to young and old stellar populations in the simulated galaxy. 
This is a particularly interesting quantity for real galaxies since with this knowledge we can
relate the power radiated by grains to the star forming activity on a range of physical scales.  

\begin{figure*}
\includegraphics[scale=0.3]{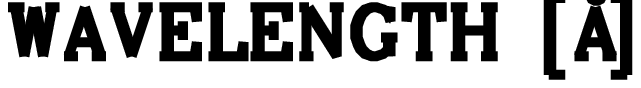}
\caption{Total intrinsic stellar emission SED (black curve) of the simulated galaxy and contributions due to young (blue) and old (red)  
stellar populations.}  
\label{fig_stellar_em_sed}
\end{figure*}

In order to calculate this fraction for each cell in the 3D grid, we first performed an additional RT calculation where we included only the 
young stellar populations (SPs). For this calculation, we derived the stellar volume emissivity in each cell by considering only the stellar 
particles with ages younger than $t_{\rm age}=1.5\times10^{8}$\,yr. The spatially integrated intrinsic emission SED of the young SPs 
so defined are shown in Fig.\ref{fig_stellar_em_sed}, together with the emission SEDs of the old SPs (due to stellar particles with age higher 
than $t_{\rm age}$) and of the total SPs. The young and old SP emission SEDs dominate in the UV and in the optical, respectively, with a 
transition region at about 3000\,\AA{}. By integrating the SED spectra over 
wavelength, we find that the young SP emission luminosity $L_{\rm young}^{\rm stellar}$ accounts for 37\% of the total stellar emission.\\
After performing the RT calculation including only young SPs, we used the procedure described in section \ref{sect_calc_y_frac} to derive 
 both the bolometric fraction of dust emission powered by young SPs for each cell, 
$L_{\rm young}^{\rm cell, abs}/L_{\rm tot}^{\rm cell,abs}$  
(see Eqs. 3 and 5) and the corresponding monochromatic fraction, $\Gamma_{\rm \lambda,young}^{\rm cell}$ (see Eq. 9).
By summing the different contributions to the dust emission over all cells, we then derived the total dust emission powered by either young or 
old SPs at each infrared wavelength ($\Gamma_{\rm \lambda,young}$ and $\Gamma_{\rm \lambda,old}$ respectively) as well as the total bolometric 
contributions ($L_{\rm young}^{abs}$ and $L_{\rm old}^{\rm abs}$).

\begin{figure*}
\includegraphics[scale=0.3]{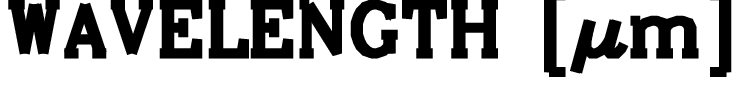}
\includegraphics[scale=0.3]{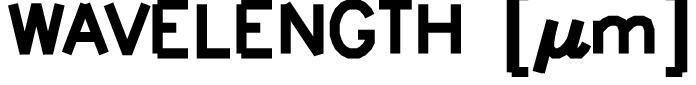}
\caption{Total dust emission SED (black curve) of the simulated galaxy and contribution due to dust emission powered by young stellar populations 
(dashed curve).}  
\label{fig_dust_em_sed}
\end{figure*}

The contribution from young and old SPs to the total dust emission can be seen in the left panel of Fig.\ref{fig_dust_em_sed}, where we show 
the integrated dust emission SED of the simulated galaxy together with the contributions powered by young and old stellar populations. 
We find that young SPs contribute to 54\% of the dust emission integrated over wavelength, that is, only slightly more than 
 old SPs.
However, we note that the fraction of young SP emission absorbed by dust is $f_{\rm young}^{\rm abs}=
\frac{L_{\rm young}^{\rm abs}}{L_{\rm young}^{\rm stellar}}=52\%$ while the fraction of old 
SP emission absorbed by dust is $f_{\rm old}^{\rm abs}= \frac{L_{\rm old}^{\rm abs}}{L_{\rm old}^{\rm stellar}}=26\%$. As expected, the 
young SP are substantially more obscured by dust than the old SPs.
The right panel of Fig.\ref{fig_dust_em_sed} shows the relative fraction of dust emission powered by young SPs at each wavelength. 
Interestingly, we notice that at wavelengths 
shorter than 60$\mu m$ the dust emission is dominated by young SP heating. This is due to several factors: the stochastical heating of the 
small grains which have higher cross sections for UV photons and tend to emit in the MIR; the contribution to the MIR from hot dust close to very 
young SPs; the MIR line emission from PAH molecules predominantly powered by UV photons. For wavelengths longer than about 60$\mu m$ the old 
SPs provide more heating than the young SPs, which nevertheless still contribute about 40-45\% of the total FIR/submm emission.\\  
Although the dust content and distribution of the simulated galaxy has some peculiar characteristics compared to that typically 
observed in nearby normal spiral galaxies (see Section \ref{res_sect_intrinsic}), the above results provide qualitative hints about the 
reliability of infrared star formation rate (SFR) indicators. These tracers can be considered robust if the majority of the dust emission 
is powered by young SPs and if the dust emission luminosity is proportional to the young SP luminosity. From the values given above 
one can realize that, although the young SPs power only half of the bolometric dust emission, the additional heating from old SPs is such that 
the total dust luminosity is approximately equal to the total young SP luminosity (within 4\%). Therefore, even if the young SPs are not 
completely obscured, measuring the bolometric dust luminosity of this galaxy would be a good way to derive the intrinsic young SP luminosity 
and thus the global SFR. This however may simply be a coincidence since galaxies are known to have varying relative luminosities of old 
and young SP according to the galaxy mass. So, it is possible that the infrared luminosity would have a different relation to the SFR for 
galaxies at different mass. Concerning the reliability of monochromatic infrared tracers of SFR, we notice from our simulation that the fraction of heating due to 
young SPs varies substantially over the infrared wavelength range (as shown in Fig.\ref{fig_dust_em_sed}). In principle, the mid-infrared 
range (1-30$\mu$m) seems to be the most appropriate for measuring galaxy SFRs, since the majority of dust emission is powered by young SPs 
in that range. However, we note that the calibration of SFR tracers at wavelengths coinciding with PAH line emission might be difficult because of the known spread in PAH 
abundances among galaxies (see e.g. Draine  et al. 2007). On the other hand,
our results bring clear support for the use of SFR tracers based on the 24$\mu$m luminosity (e.g. Wu et al. 2005, Rieke et al. 2009). 
In the range 30-100$\mu$m, the fraction of young SP heating varies quickly. This makes the use of 
monochromatic SFR tracer in this range quite problematic. Finally, at wavelengths longer than 100$\mu$m, our results suggest that 
SFR tracers cannot be based on the assumption that the majority of dust heating is contributed by young SPs.

\begin{figure*}
\includegraphics[scale=0.47]{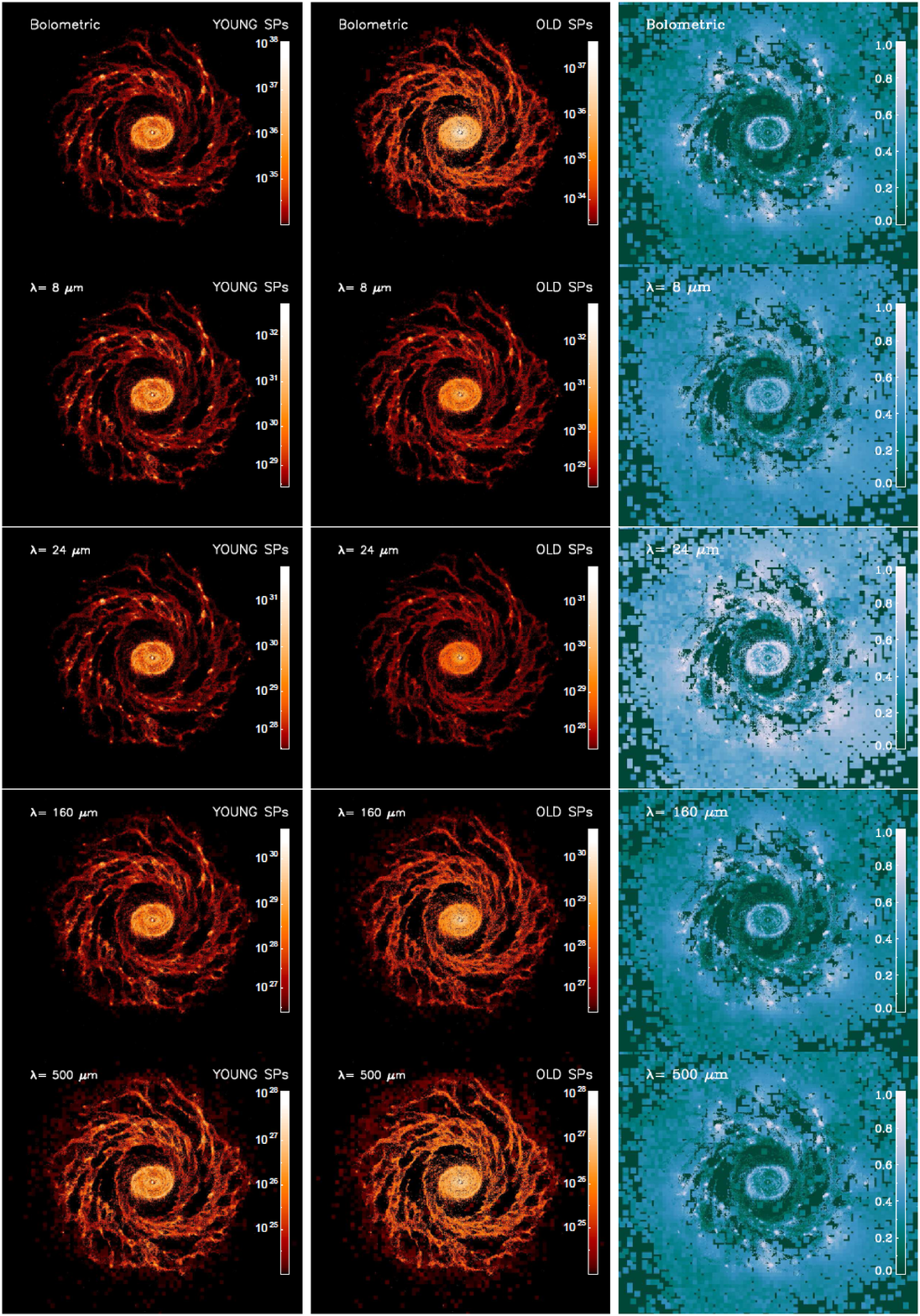}
\caption{Bolometric and monochromatic dust emission maps powered by young SPs (left) and old SP (middle) and relative fraction of dust emission
powered by young SPs (right) for the face-on view.} 
\label{galaxy_dust_idir0_rel}
\end{figure*}

In Fig.\ref{galaxy_dust_idir0_rel} we show maps of the dust emission powered by young SPs, old SPs as well as maps of the relative fraction 
powered by young SPs for the face-on view. Interesting features we notice 
are the following: \\
(i) for the bolometric dust emission maps, the morphology tends to be more clumpy for the dust emission powered 
by young SPs compared to that powered by old SPs;\\ 
(ii) young SPs tend to dominate the heating in some discrete peaks distributed throughout the disc and in the external ring around the nuclear
disc; \\
(iii) as expected, the 24$\mu$m map shows generally a larger fraction of emission powered by young SPs at every position compared to the other maps; \\

\begin{figure*}
\includegraphics[scale=0.5]{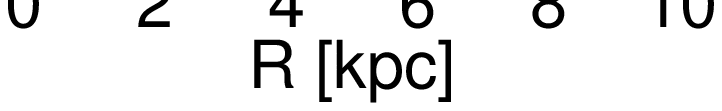}
\caption{Radial average of the young SP powered dust luminosity fraction (red curve) and young SP powered dust luminosity radial density (black curve) 
profiles for the face-on view.}  
\label{prof_young_idir0}
\end{figure*}

In the panels of Fig.\ref{prof_young_idir0} 
we show the radial average profile for the relative fraction of the 
young SP powered dust emission and the radial 
density profile for the young SP powered dust luminosity. We notice that:\\
(i) the peak of the young SP powered dust luminosity density is at the exterior part of the nuclear disc;  \\
(ii) in the nuclear disc the relative fraction of young SP powered dust emission peaks at the external boundary for both the bolometric map 
and the monochromatic maps. However, the variation between inner and outer parts of the nuclear disc is smoother for the 8 and 
24 $\mu$m maps. \\
(iii) the spiral disc shows a young SP powered dust emission fraction profile which is oscillating around 40-50\% except for the 
24$\mu$m case where the average is approximately 70\%. 
From the cumulative fraction radial profile of the young SP powered dust luminosity, shown in Fig. D1 in the appendix, 
we notice that most ($>$60\%) of the young SP powered dust luminosity is emitted by the nuclear disc for both the bolometric and all the 
other maps, except the 500$\mu$m map where the fraction goes down to 50\% \\
(iv) at radii larger than 6\,kpc, the heating of the dust is mainly provided by old SPs (except at 24\,$\mu$m).

For the maps and radial profiles derived for the inclined view, shown in Fig. D2-D4 of 
the appendix, we notice essentially the same characteristics for all the maps and radial profiles as for the face-on view.\\

\begin{figure*}
\includegraphics[scale=0.47]{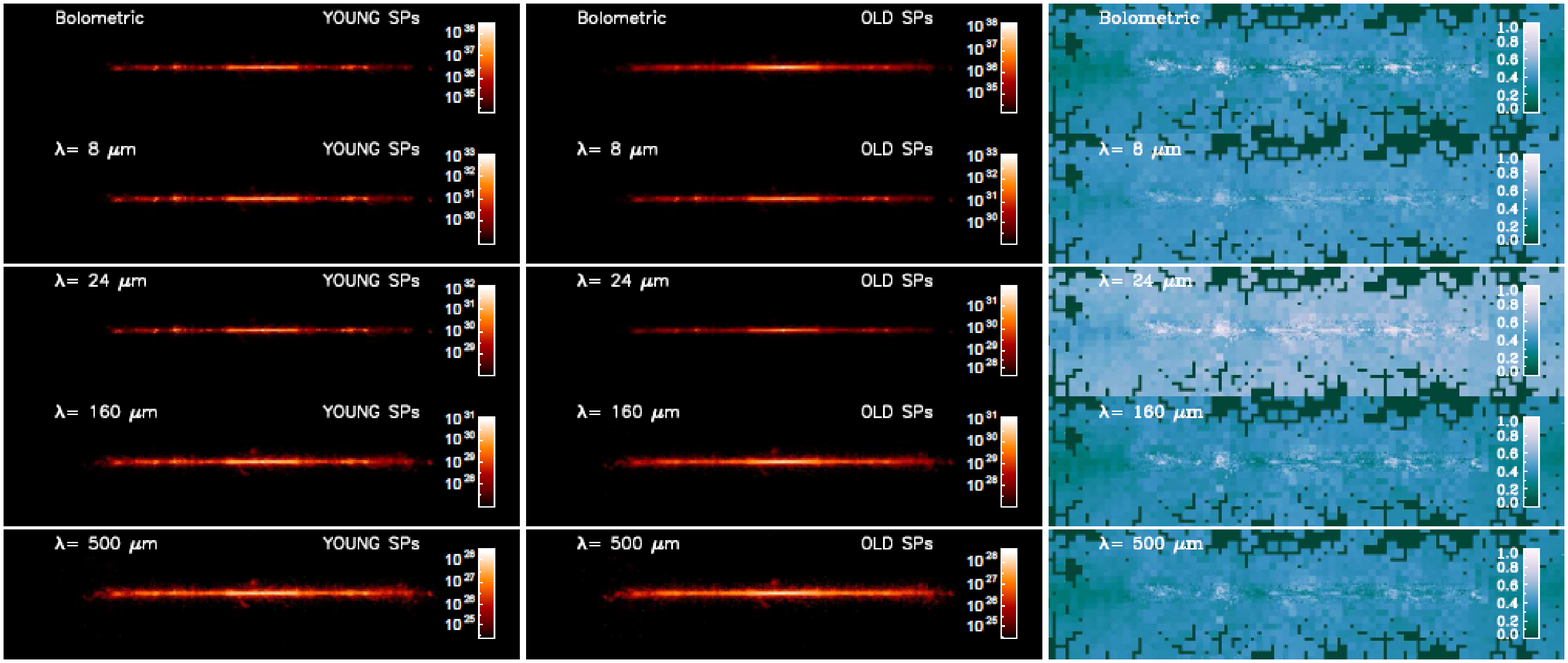}
\caption{Bolometric and monochromatic dust emission maps powered by young SPs (left) and old SP (middle) and relative fraction of dust emission
powered by young SPs (right) for the edge-on view} 
\label{galaxy_dust_idir7_rel}
\end{figure*}

Finally, in Fig.\ref{galaxy_dust_idir7_rel}
we show the young and old SPs powered dust emission maps and the relative fraction of emission due to young SPs for the edge-on view. 
We notice that:\\
(i) the dust emission powered by young SPs is rather more clumpy within the galaxy disc than that powered by old SPs;\\ 
(ii) as for the face-on and inclined views, the fraction is higher for the 24$\mu$m emission compared to the other wavelengths and the 
bolometric emission.\\

As for the other inclinations, we derived the profiles of the vertical average of the young SP powered dust emission 
fraction and the young SP powered dust luminosity vertical density and cumulative fraction (see Fig. D5-D6 in the appendix). 
From these we see that more than 95\% of the young SP powered dust emission luminosity is emitted within 100\,pc of the galaxy plane. The fraction of young SP powered 
dust emission is about 50\% in the galaxy plane and goes down to 40\% at z=1\,kpc, with the exception of 
the 24$\mu$m map where the fraction goes from 75\% to 60\%. 

\section{Summary}
In this paper we have extended the capabilities of the 3D ray-tracing RT code \textsc{DART-Ray} to include a self-consistent 
calculation of stochastically heated dust emission, using the same algorithms applied to the axi-symmetric models of Popescu et al. (2000)
and Popescu et al. (2011). This calculation can be performed either by using a brute force approach (where 
the derivation of the non equilibrium dust emission spectra is performed exactly for each cell) or by using an adaptive SED library approach. 
In the latter case, dust emission spectra are derived for the average RFED of the cells having similar wavelength--integrated UV and optical 
RFEDs. The spectra so calculated are then assigned to the corresponding cells with similar RFEDs.  

We also gave a general prescription for the calculation of the fraction of dust emission powered by different stellar populations in a galaxy. 
This prescription is in principle applicable to any radiation transfer code and any galaxy model/simulation. It was already used in the 
calculations from Popescu et al. (2000), although the method was never described before.

To show the full capabilities of the code, we used \textsc{DART-Ray} to derive the stellar and non equilibrium dust emission maps of a 
high-resolution N-body+SPH galaxy simulation for different wavelengths and inclinations. The derived maps have been used to perform an 
analysis of the dust attenuation, scattering and heating properties of the simulated galaxy. 
This analysis has confirmed several features which have been previously quantitatively investigated only with axi-symmetric models.
In particular this analysis has highlighted:\\
(i) the importance of dust in modifying the radial and vertical profiles of the perceived stellar emission of the simulated galaxy from those 
of their intrinsic emission;\\
(ii) the flattening of the attenuation curve, compared to the extinction curve of the assumed dust model, at high galaxy inclinations. 
This effect can be explained in terms of the relative geometry of the stellar and dust distribution without the need of a modified dust model;\\
(iii) the presence of ``negative attenuation'' on local scales: the observed stellar emission at certain positions within the simulated 
galaxy is enhanced compared to the emission that would be observed in the absence of dust;\\
(iv) in the edge-on view, the dust lane obscuring the stars close to the galaxy plane is seen only at optical/near-IR wavelengths but not 
at UV wavelengths;\\
(v) the fraction of the total observed emission due to scattered light is less than 20\% at all wavelengths. However, higher fractions of 
scattered radiation are reached at local positions within and outside the galaxy disc;\\
(vi) young and old SPs power approximately the same amount of total dust emission luminosity, with the young SPs providing slightly more heating 
(54\%). However, the total dust emission is powered predominantly by young stars for $\lambda<60\mu m$ and by old stars for $\lambda>60\mu m$;\\
(vii) the dust emission powered by young SPs present a more clumpy morphology compared to that powered by old SPs. This is due to the 
predominance of young SPs in powering the dust emission in the vicinity of young star formation regions. We note that UV photons also propagate
at large distances from the star formation regions and contribute substantially to the heating of the diffusely distributed dust at 
all wavelengths.

\section*{Acknowledgements}
G.N. and C.C.P. would like to acknowledge support from the Leverhulme Trust research project grant RPG-2013-418. 
V.P.D. is supported by STFC Consolidated grant \# ST/J001341/1. The galaxy simulation used in
this study was run at the High Performance Computer Facility of the University of Central Lancashire.  
We would like to thank the referee for his/her detailed and comprehensive report, which helped improve the paper.


\appendix

\section*{APPENDIX}
Additional figures and plots for section \ref{results_section} can be found in the appendix available in the on-line version of the paper.

\end{document}